\documentclass[useAMS,usenatbib]{mn2e}

\usepackage[psamsfonts]{amssymb}
\usepackage[dvips]{graphicx}
\usepackage{amsmath,alltt}
\usepackage{multirow}
\usepackage{rotating}
\usepackage{lscape}

\usepackage{enumerate,multirow}
\usepackage{amsmath}
\usepackage{amssymb} 
\usepackage{color}
\usepackage{wasysym}

\def\gap{\;\rlap{\lower 2.5pt
    \hbox{$\sim$}}\raise 1.5pt\hbox{$>$}\;}
\def\lap{\;\rlap{\lower 2.5pt
   \hbox{$\sim$}}\raise 1.5pt\hbox{$<$}\;}
\def\beq{\begin{equation}}
\def\eeq{\end{equation}}

\def\msun{M_\odot}

 \usepackage{ulem}

\title[Enigmatic Stellar Populations in Galactic Nuclei]{On the origins of enigmatic stellar populations in Local Group galactic nuclei}
\author[Leigh, Antonini, Stone, Shara \& Merritt]{Nathan W. C. Leigh$^{1}$, Fabio Antonini$^{2}$, Nicholas C. Stone$^{3,4}$, Michael M. Shara$^{1}$, 
\newauthor
David Merritt$^{5}$
\thanks{E-mail: nleigh@amnh.org (NL); fabio.antonini@northwestern.edu (FA); nstone@phys.columbia.edu (NS); mshara@amnh.org (MS); tboeker@cosmos.esa.int (TB); merritt@astro.rit.edu (DM)}\\
$^{1}$Department of Astrophysics, American Museum of Natural History, Central Park West and 79th Street, New York, NY 10024 \\
$^{2}$Center for Interdisciplinary Exploration and Research in Astrophysics (CIERA) and Department of Physics and Astronomy, \\ 
Northwestern University, 2145 Sheridan Rd, Evanston, IL 60208, USA \\
$^{3}$Columbia Astrophysics Laboratory, Columbia University, New York, NY, 10027, USA \\
$^{4}$Einstein Fellow\\
$^{5}$Department of Physics and Center for Computational Relativity and Gravitation, Rochester Institute of Technology, Rochester, NY 14623, USA}

\begin{document}

\pagerange{\pageref{firstpage}--\pageref{lastpage}} \pubyear{2016}

\maketitle

\label{firstpage}

\begin{abstract}

We consider the origins of enigmatic stellar populations in four Local Group galactic nuclei, specifically the Milky Way, M31, M32 and M33.  These are centrally concentrated blue stars, found in three out of the four nuclear star clusters (NSCs) considered here.  Their origins are unknown, but could include blue straggler (BS) stars, extended horizontal branch stars and young recently formed stars.  Here, we calculate order-of-magnitude estimates for various collision rates, as a function of the host NSC environment and distance from the cluster centre.  These rates are sufficiently high that BSs, formed via collisions between main sequence (MS) stars, could contribute non-negligibly ($\sim$ 1-10\% in mass) to every surface brightness profile, with the exception of the Milky Way.  Stellar evolution models show that the envelopes of red giant branch (RGB) stars must be nearly completely stripped to significantly affect their photometric appearance, which requires multiple collisions.  Hence, the collision rates for individual RGB stars are only sufficiently high in the inner $\lesssim$ 0.1 pc of M31 and M32 for RGB destruction to occur.  Collisions between white dwarfs and MS stars, which should ablate the stars, could offer a steady and significant supply of gas in every NSC in our sample.  The gas could either fragment to form new stars, or accrete onto old MS stars already present.  Thus, collisional processes could contribute significantly to the observed blue excesses in M31 and M33; future studies should be aimed at better constraining theoretical predictions to compliment existing and future observational data.
\end{abstract}

\begin{keywords}
galaxies: nuclei -- blue stragglers - galaxies: photometry -- stars: kinematics and dynamics.
\end{keywords}

\section{Introduction} \label{intro}

Over the last few decades, observational surveys have achieved unprecedented resolution, reaching 
sufficient sensitivity to resolve the central parsec of not only the Milky Way (MW), but also its nearest 
neighbors in the Local Group, namely M31, M32 and M33.  These observations revealed a number 
of anomalous stellar populations, typically very blue and massive stars whose presence is 
difficult to reconcile without an in situ origin.\footnote{This is because the mechanisms invoked to migrate 
blue stars inward after forming elsewhere in the Galaxy tend to operate on characteristic timescales 
that exceed the (inferred) mean ages of the stars.}  The origins of these curious populations have 
become an active area of research over the last few years.  

Within the core of the MW nuclear star cluster (NSC) complex, about a hundred OB stars reside in a strongly 
warped disk (between 0.8 and 12 arcsec, or $\sim$ 0.04-0.6 pc from the SMBH) \citep[e.g.][]{baruteau11,chen15}, rotating 
clockwise on the sky.  The mean 
stellar mass and age are inferred to be $\gtrsim$ 10 M$_{\odot}$ and $\sim$ 6 Myr, respectively.  There exists 
tentative evidence for a second, less-massive and counterclockwise disk that consists of stars similar 
in their photometric, spectroscopic and even kinematic properties \citep[e.g.][]{paumard06,bartko09,bartko10}.  
Closer in to the very centre of the nucleus is an even more centrally ($\lesssim$ 0.01 pc) compact structure 
called the S-star cluster consisting of main-sequence B-stars with randomly oriented and eccentric orbits 
\citep{bartko10}, with eccentricities ranging up to 0.8 and having a median value of $\sim$ 0.4 
\citep[e.g.][]{lu09,bartko09,bartko10}. 

The existence of \textit{young} stars in the very inner regions of the Galactic Centre would be puzzling, given that a 
super-massive black hole (SMBH) is also present, called Sgr A*.  The standard picture for 
star formation involving the collapse of a self-gravitating cold molecular cloud is unlikely to happen due to 
the strong tidal field of the SMBH.  But the physics underlying star formation in such an extreme environment 
close to an SMBH is highly uncertain, and other mechanisms can potentially be invoked \citep[e.g.][]{alexander06}.  
For example, \citet{bonnell08} showed that, when a molecular cloud passes close to an SMBH, it can become 
tidally disrupted and form a thin accretion disk.  Star formation could then occur if the density in the disk is 
sufficiently high for its own self-gravity to overcome the tidal force from the SMBH \citep[e.g.][]{levin03,nayakshin05}.  
Alternatively, a dense young cluster could migrate inward due to dynamical friction if it is sufficiently massive 
($\gtrsim$ 10$^5$ M$_{\odot}$), provided it originates within a few parsecs of the central SMBH 
\citep[e.g.][]{gerhard01}.  

Another mechanism to explain the blue stars at the Galactic Centre is stellar mergers, which can form 
rejuvenated main-sequence stars that appear young \citep[e.g.][]{antonini11,ghez05}.  This could be facilitated 
by the central SMBH, which forms an outer triple companion to any binary orbiting sufficiently close to it 
that its orbit is Keplerian \citep[e.g.][]{antonini10}.  Thus, stellar binaries could undergo large oscillations in their eccentricities 
and inclinations due to the Kozai-Lidov mechanism (for a review, see \citet{naoz16}).  Tidal damping 
between binary companions could then shrink or circularize their orbits, or even drive a direct collision 
if the eccentricity becomes sufficiently high.  Merger products could also form when one of the binary 
components evolves off the main-sequence to over-fill its Roche lobe \citep{prodan15,stephan16}. 

In addition to the surplus of blue stars very close to the central SMBH, observations suggest 
a relative paucity of old evolved stars in the central half-parsec of our Galaxy 
\citep[e.g.][]{krabbe91,najarro94,buchholz09,do09,bartko10}.  Because of the extreme extinction 
in the direction of Sgr A*, these surveys reached a limiting magnitude of K$_{\rm s} \sim$ 18, nevertheless 
indicating that the red giant branch (RGB) and horizontal branch (HB) stars with ages $\gtrsim 1$ Gyr and 
masses 0.5 - 4 M$_{\odot}$ are less numerous than expected when assuming a normal (e.g. Kroupa) 
initial stellar mass function \citep{genzel10}.     

Several mechanisms have been suggested to explain the missing RGB stars in the Galactic Centre, 
including star-star collisions \citep[e.g.][]{genzel96,davies98,bailey99,dale09}, three-body interactions 
with a central SMBH-SMBH binary \citep[e.g.][]{baumgardt06,portegieszwart06,gualandris12} and the 
infall of star clusters \citep[e.g.][]{kim03,antonini12}.  Motivated by observations of the disk 
of blue stars orbiting Sgr A* at $\sim$ 0.04 - 0.6 pc, \citet{amaroseoane14} recently proposed 
star-disk collisions as a mechanism to deplete RGB stars which, in this scenario, collide repeatedly 
with star-forming cores in the natal gaseous disk that is speculated to form the blue stars.  

In the nearby spiral galaxy M31, our current understanding of the inner nucleus is slightly less detailed 
than in the MW.  This is in spite of the greater extinction in the direction of the Galactic Centre, and is partly 
due to the larger distance to M31, making it less well resolved \citep[e.g.][]{light74}.  M31 is home to two 
nuclear disks.  A massive old disk of stars, called the P1+P2 complex, constitutes the inner nucleus,
with a total mass of $\sim$ 3 $\times$ 10$^7$ M$_{\rm \odot}$ and a half-power radius of
 $\sim$ 1.8 pc \citep{bender05}.  Inside this old disk and immediately surrounding the 
central SMBH in M31 is a compact nuclear cluster of blue stars 
dubbed P3, organized into a disk with half-power radius $\sim$ 0.2 pc \citep{lauer98}.  The 
spectrum for P3 is consistent with a clump of stars in the spectral range B5-A5, with 
an integrated magnitude M$_{\rm V} =$ -5.7 \citep{lauer98,bender05,demarque07}.\footnote{\citet{bender05} 
created a mock colour-magnitude diagram for P3 composed of 186 stars of types A5 to B5 plus 7 evolved
red giants, which can reproduce the observed spectrum.  From this, the authors derived a total mass
of $\sim$ 4200 M$_{\odot}$ (assuming many additional unseen low-mass main-sequence stars).} 

Many hypotheses have been proposed to explain the blue cluster at the very centre 
of M31.  Several authors have invoked a recent burst of star formation near the 
central SMBH to explain these anomalous blue stars.  For example, \citet{bender05} 
illustrated that the integrated spectrum and spectral energy distribution is consistent 
with a burst of star formation $\sim$ 200 Myr ago.  This was later improved upon 
by \citet{lauer12}, who fit Padova isochrones for single-burst populations with ages of 
50, 100 and 200 Myr to an observed colour-magnitude diagram of the blue 
cluster P3.  \citet{chang07} argued that gas from stellar evolution-induced mass loss can 
reach surface densities sufficiently high to induce fragmentation and star formation 
in the immediate vicinity of the SMBH.  In this picture, the gas ends up being funneled 
into orbit about the SMBH due to the non-axisymmetric structure of the surrounding M31 
nucleus \citep[e.g.][]{levin03,bonnell08,wardle12}. 

Alternatively, \citet{demarque07} pointed out that the observed spectrum of the hot 
stars in P3 is also consistent with an 
old stellar population in the extended or post-horizontal branch phase of evolution, or even 
blue stragglers \citep[e.g.][]{sills97,sills01}.  The blue stragglers (BSs) could 
be created from direct stellar collisions and/or mergers, while the extended 
horizontal branch (EHB) stars could be the result of tidal stripping of red giant 
branch stars by the SMBH \citep[e.g.][]{genzel96,davies98}.  \citet{yu03} 
considered a collisional origin for the hot stars in M31.  The author found a 
collision rate that is too small to account for their origins, however this 
result rests on the critical assumption that the collisions occur within a 
spherical stellar distribution surrounding the central SMBH.  More recent 
observations have demonstrated that this assumption is not correct, and the 
distribution more closely resembles the eccentric disk model of \citet{tremaine95} (see 
Section~\ref{app} for more details).  

In the nearby dwarf elliptical M32, there is no evidence for a central blue excess in the inner nucleus.  
There is little indication of any change in the optical and UV colours in the very inner 
regions of the nucleus close to the putative central SMBH.  Although slightly bluer colors are observed 
in the very central 
pixels of a WFPC F555W-F814W color image by \citet{lauer98}, the U-V colours 
should more sensitively probe for a central blue excess, and this becomes \textit{redder} toward 
the centre of the galaxy.  The authors conclude that M32 does \textit{not} harbor an excess 
of blue stars in the very inner regions of the nucleus.  Weaker CO lines and redder 
continuum are also observed in the inner $\sim$ 0.9 pc, but this is thought to be due to dust 
emission.  This trend cannot be explained from changes in the stellar populations in the M32 
nucleus, unlike the MW NSC where weaker CO lines are attributed to a dearth of RGB stars 
\citep{sellgren90}.  The stellar density in M32 is one of the highest ever observed 
($\sim$ 10$^7$ M$_{\odot}$), and should (naively) be more than 
sufficient for collisions to occur.  However, missing RGB stars cannot by themselves explain the observed 
trend between weaker CO lines and a redder continuum reported by \citet{seth10}, since this 
would contribute to a bluer, not redder, K-band continuum.

A number of mechanisms have been proposed for the origins of the M32 nucleus, which consists 
of a puffed up disk embedded in a more spherical stellar distribution.  \citet{seth10} suggest 
that the disk component could have formed by gas accretion into the nucleus, triggered by a major or minor 
merger in the recent past.  The author concedes, however, that secular processes alone could 
still be responsible for the disk kinematics.  Another possibility for the formation of the nuclear 
disk in M32, originally proposed by \citet{bailey80} and later (briefly) re-visited by \citet{seth10}, 
is the gradual build-up of gas from stellar winds in the surrounding population.  This might 
naturally predict a trend in stellar age as a function of radius and the observed lack of any break in 
the rotation curve, while also potentially accounting for an observed abundance gradient in 
which the [Mg/Fe] ratio begins at subsolar values in the inner nucleus and increases at larger radii 
\citep{worthey04,rose05}.  

Finally, in the nearby spiral galaxy M33, a strong B-R colour gradient exists in the nucleus, such that the 
central part of the NSC is bluer than its outskirts.  \citet{schmidt90} tried to reproduce the spectrum of the 
M33 nucleus using a population synthesis method 
comprised of a series of star clusters of different ages.  As further discussed in \citet{vandenbergh91}, the results 
find that the integrated spectrum can be explained if it is dominated by a young metal-rich population, which is 
very different from the spectrum of an old 
globular cluster in the near-infrared.  Using high resolution photometry, \citet{kormendy93} and later \citet{lauer98} 
noted a colour gradient in the optical, with a central concentration of blue stars.  This was later confirmed via 
high resolution UV imaging; the nucleus is bluer and even more compact than at longer wavelengths (see 
also \citet{carson15}).  Surface photometry of the nuclear region 
in M33 suggests an average isophotal shape $\epsilon =$ 0.16 $\pm$ 0.01, with the NSC becoming 
rounder toward its centre.  Given that the relaxation time in M33 is thought to be very short, this ellipticity 
is thought to be due to rotation instead of an anisotropic velocity ellipsoid \citep{lauer98}.  

Both \citet{kormendy93} and \citet{lauer98} discuss this 
central blue excess in the context of blue stragglers (BSs) formed from stellar collisions.  \citet{kormendy93} 
further suggested that the short relaxation time could mean that core-collapse has previously occurred, which 
might increase the collision rate in the inner nucleus.  Indeed, the half-power 
radius of the central nuclear region is only 0.13 pc.  \citet{long02} 
performed stellar population modeling of STIS spectra, and found that the data can best be fit assuming 
two distinct stellar populations, with masses and ages of 9000 M$_{\odot}$ at 40 Myrs and 
7.6 $\times$ 10$^4$ M$_{\odot}$ at 1 Gyr, respectively.  

In this paper, we re-visit the issue of the origins of the blue stars observed in the very central regions of 
Local Group nuclei, with an emphasis 
on comparing and contrasting the different theoretical mechanisms proposed to explain the observed 
properties of these curious populations.  Throughout this paper, we will use the term "blue stars" to 
refer to a secondary stellar population, defined relative to an underlying primary population with a 
much redder mean colour.  The total amount of blue light is sufficient that accounting for both populations 
with a single stellar mass function would require some non-canonical spike or special feature in order to 
reproduce the secondary population.  The origins of these blue stars are unknown, as are their ages.  We 
will use the term "young stars" to refer to blue stars that are very young relative to a primary old 
population.  

We focus on the different channels for mediating direct collisions and mergers (involving old stars), 
as a function of the host nuclear star cluster environment.  In Section~\ref{rates}, 
we present order-of-magnitude estimates for the rates of MS+MS, MS+RGB, MS+WD, WD+RGB, 
MS+NS, BH+RGB and single-binary (1+2) collisions, suitable for different assumptions about 
the properties of the nucleus.  In Section~\ref{ehbform}, we present stellar evolution models of tidally stripped 
RGB stars, which illustrate that the photometric appearance and subsequent evolution are roughly insensitive 
to the degree of mass loss (unless nearly the entire envelope is removed).  
We discuss in Section~\ref{app} the observed properties of the nuclei in the MW, M31, M32 and M33, which 
decide the appropriate form of our derived collision rates, and we present 
the calculated collision rates for these different NSCs, as a function of the distance from the cluster centre.  
We discuss in Section~\ref{discussion} the implications of our results for the role of collisional 
processes in shaping the observed properties of dense unresolved stellar populations.  We go on to 
discuss the relevance of our results for inferring age spreads in nuclear star clusters, as well as the most 
probable formation mechanism for the inner blue stars observed in three out of four NSCs in the 
Local Group.  We summarize our main conclusions in Section~\ref{summary}.  

\section{Collisions and mergers in nuclear star clusters} \label{collisions}

In this section, we derive order-of-magnitude estimates for the mean rate of collisions and mergers, as a function 
of the host nuclear cluster environment.  These include direct collisions between 
main-sequence (MS) stars, tidal stripping of red giant branch (RGB) stars via direct MS+RGB, WD+RGB 
and BH+RGB collisions, direct collisions between main sequence stars and both neutron stars (NSs) 
and white dwarfs (WDs) (which 
could serve as a prolific source of gas in the nucleus; see Section~\ref{discussion} for more details), 
direct encounters between stellar binaries and MS stars (which typically produce a MS+MS collision) and 
Kozai-Lidov oscillations mediated by an SMBH acting as the outer third companion.  All collisions 
and mergers are assumed to involve only old stars, in an effort to evaluate if each mechanism could 
account for the blue populations observed in three out of the four nuclei considered here.  Note that dense 
galactic nuclei have high velocity dispersions, and are hence predicted to have low binary fractions 
\citep[e.g.][]{heggie75,leigh13,leigh15}.  Thus, 
binary-binary (2+2) encounters are irrelevant for our purposes, since 1+2 encounters dominate over 2+2 
encounters for binary fractions $\lesssim$ 10\% \citep{sigurdsson93,leigh11}.

\subsection{Collision rates} \label{rates}

First, we calculate the mean rate of collisions in a nuclear star cluster with a central 
SMBH of mass M$_{\rm BH}$.  We begin by considering a nuclear disk with predominantly 
Keplerian orbits about the central SMBH.  We then go on to consider a spherical pressure-supported 
system surrounding a central SMBH.  We then combine these limiting cases in order to provide a 
single formula for the collision rate that is applicable to every NSC in our sample since, 
as described in Section~\ref{app}, each NSC can be reasonably well described by one of these two 
limiting regimes. 

The stellar orbits surrounding a central SMBH will be roughly Keplerian if they are within the 
radius of influence r$_{\rm inf}$ of the SMBH.  For a nuclear disk, for example, the bulk of the stellar orbits will 
be approximately Keplerian provided the influence radius significantly exceeds the scale length of the disk a.  
The influence radius is approximately defined as \citep{merritt13}: 
\begin{equation}
\label{eqn:rinf}
r_{\rm inf} = \frac{GM_{\rm BH}}{\sigma^2}, 
\end{equation}
where $\sigma$ is the velocity dispersion of the surrounding NSC.

We begin by calculating and comparing the relevant velocities, namely the 
escape velocity from the stellar surface v$_{\rm e}$, the Hill velocity v$_{\rm H}$ 
and the velocity dispersion $\sigma$.  A comparison of these velocities determines the importance 
of gravitational-focusing during collisions.  We assume that all stars 
in the NSC are identical, with masses M and radii R.  For (old) low-mass main-sequence 
stars, a reasonable approximation is M/R $\sim$ M$_{\odot}$/R$_{\odot}$.  Thus, the escape 
velocity from the surface of a typical star in the disk/cluster is:
\begin{equation}
\label{eqn:escape}
v_{\rm e} = \Big( \frac{2GM}{R} \Big)^{1/2},
\end{equation}
which gives $\sim$ 618 km/s assuming M $=$ 1 M$_{\odot}$ and R $=$ 1 R$_{\odot}$.

At the Hill radius R$_{\rm H}$, the orbital frequency of a body in orbit about a star is 
comparable to the orbital frequency of the star around the central SMBH.  The Hill velocity 
is then the characteristic velocity associated with the Hill radius, or:
\begin{equation}
\label{eqn:vHill}
v_{\rm H} \approx \Big( \frac{GM}{R_{\rm H}} \Big)^{1/2}, 
\end{equation}
where
\begin{equation}
\label{eqn:rHill}
R_{\rm H} \approx r\Big( \frac{M}{M_{\rm BH}} \Big)^{1/3},
\end{equation}
and r is the distance from the central SMBH.  Plugging Equation~\ref{eqn:rHill} 
into Equation~\ref{eqn:vHill}, we obtain:
\begin{equation}
\label{eqn:vHill2}
v_{\rm H} \approx \Big( \frac{GM^{2/3}M_{\rm BH}^{1/3}}{r} \Big)^{1/2}. 
\end{equation}
In M31, for example, Equation~\ref{eqn:vHill2} yields v$_{\rm H}$ $\sim$ 1.1 km/s, assuming a scale length for 
the disk of a $=$ 1.8 pc \citep{bender05}, M $=$ 1 M$_{\odot}$ and M$_{\rm BH} =$ 1.4 $\times$ 10$^8$ M$_{\odot}$ \citep{bender05}.  

If $\sigma$ $>$ v$_{\rm e}$, then we are in the super-escape regime, so that gravitational-focusing 
can be neglected.  Conversely, in the sub-escape and super-Hill regime v$_{\rm H} <$ $\sigma$ $<$ v$_{\rm e}$ 
gravitational-focusing is significant.   We do not consider the sub-Hill regime (i.e. $\sigma <$ v$_{\rm H}$), 
since the Hill velocity is always very small in galactic nuclei.  

Putting this all together, the collision rate for a Keplerian nuclear disk can be written:  
\begin{equation}
\label{eqn:gamma1}
\Gamma \approx \Big( \frac{{\Sigma}{\Omega}R^2}{M} \Big)\Big(1 + \Big( \frac{v_{\rm e}}{\sigma} \Big)^2 \Big),
\end{equation}  
where 
\begin{equation}
\label{eqn:omega}
\Omega = \Big( \frac{GM_{\rm BH}}{r^3} \Big)^{1/2},
\end{equation}
and the second term in Equation~\ref{eqn:gamma1} smoothly transitions the collision rate into and out of the regime 
where gravitational-focusing becomes important.  The stellar surface mass density is:
\begin{equation}
\label{eqn:surfdens}
\Sigma = \Big( \frac{M_{\rm d}}{\pi{a^2}} \Big),
\end{equation}
where M$_{\rm d}$ is the total stellar mass of the disk, a is its scale length and h is its scale height.  
Equation~\ref{eqn:gamma1} gives the mean rate at which a given star or object orbiting at a distance r 
from the central SMBH collides with other objects.  

Next, we consider an NSC for which the dominant stellar distribution is (approximately) spherical, pressure-supported 
and the stellar motions are predominantly random.  Here, 
we assume that the stellar orbits are distributed isotropically and the velocity distribution is isothermal (i.e., Maxwellian with 
dispersion $\sigma_{\rm 0}$) \citep[e.g.][]{merritt13}.  

Similar to Equation~\ref{eqn:gamma1}, an order-of-magnitude estimate for the mean rate $\Gamma$ at which a given star or object 
undergoes direct collisions with other objects is:
\begin{equation}
\label{eqn:gamma9}
\Gamma \approx \Big( \frac{{\rho}{\sigma}R^2}{M} \Big)\Big(1 + \Big(\frac{v_{\rm e}}{\sigma}\Big)^2\Big),
\end{equation} 
where, as before, the second term smoothly transitions the collision rate into and out of the regime where gravitational-focusing 
becomes important.  In Equation~\ref{eqn:gamma9}, we use the local velocity dispersion $\sigma$, calculated as:
\begin{equation}
\label{eqn:sigloc}
\sigma(r)^2 = \sigma_{\rm 0}^2 + \frac{GM_{\rm BH}}{r}.
\end{equation}  
This serves to modify the velocity dispersion very close to the central SMBH, where the stellar orbits begin transitioning into 
the Keplerian regime.

Finally, combining Equations~\ref{eqn:gamma1} and~\ref{eqn:gamma9}, we obtain a 
general formula for the rate of collisions in a nuclear star cluster with or without a central SMBH:
\begin{equation}
\label{eqn:gamma10}
\Gamma \approx ({\rho}\sigma + {\Sigma}\Omega)\Big( \frac{R^2}{M} \Big)\Big(1 + \Big( \frac{v_{\rm e}}{\sigma} \Big)^2 \Big),
\end{equation}
As we will show in Section~\ref{app}, Equation~\ref{eqn:gamma10} reasonably describes the collision rates in every 
galactic nucleus in our sample, provided we set \textit{either} $\rho =$ 0 or $\Sigma =$ 0.  This is because, within our chosen 
outer limit of integration (i.e., 2 pc), the dominant NSC 
structure is either well described by one of these two extremes, namely an (approximately) spherical pressure-supported 
system (i.e., the MW, M32 and M33) or a Keplerian disk (i.e., M31).

\subsection{Binary mergers due to Kozai-Lidov oscillations} \label{kozai}

In this section, we derive order-of-magnitude estimates for the mean rate of binary mergers mediated 
by Kozai-Lidov oscillations with an SMBH acting as the outer triple companion, as a function 
of the host nuclear environment and distance from the central SMBH.  

The characteristic time-scale for eccentricity oscillations to occur due to the Kozai-Lidov mechanism is \citep{holman97,antonini16}:
\begin{equation}
\label{eqn:taukl}
\tau_{\rm KL} \approx P_{\rm b}\Big( \frac{M_{\rm b}}{M_{\rm BH}} \Big)\Big( \frac{r}{a_{\rm b}} \Big)^3(1-e_{\rm BH}^2)^{3/2}, 
\end{equation} 
where M$_{\rm b} =$ 2M is the binary mass, r is the distance from the SMBH, e$_{\rm BH}$ is orbital eccentricity of the 
binary's orbit about the central SMBH, a$_{\rm b}$ is the binary orbital separation and P$_{\rm b}$ is the binary 
orbital period.  The rate at which binaries merge due to Kozai-Lidov oscillations within a given distance from the SMBH 
is then:
\begin{equation}
\label{eqn:gammakl1}
\Gamma_{\rm KL}(r) \sim \frac{{\pi}f_{\rm b}f_{\rm i}{\Sigma}r^2}{2M\tau_{\rm KL}},
\end{equation} 
for a nuclear disk, and 
\begin{equation}
\label{eqn:gammakl2}
\Gamma_{\rm KL}(r) \sim \frac{2{\pi}f_{\rm b}f_{\rm i}{\rho}r^3}{3M\tau_{\rm KL}},
\end{equation} 
for a spherical pressure-supported nucleus.  Here, f$_{\rm b}$ is the fraction 
of objects that are binaries, and 
f$_{\rm i}$ is the fraction of binaries with their inner and outer orbital planes aligned such that Kozai 
cycles operate.  For simplicity, we assume f$_{\rm i} =$ 1.  Thus, the calculated rates 
are strict upper limits, since they assume that every binary is aligned relative to the SMBH such that Kozai oscillations 
will occur, which is certainly not the case.  We will return to this issue in Section~\ref{discussion}.

Importantly, Equation~\ref{eqn:taukl}, and hence Equations~\ref{eqn:gammakl1} and~\ref{eqn:gammakl2}, 
are only valid at distances from the SMBH smaller than \citep{prodan15}:
\begin{equation}
\begin{gathered}
\label{eqn:rsc}
r_{\rm SC} \approx 0.02\Big( \frac{a_{\rm b}}{{\rm AU}} \Big)^{4/3}\Big( \frac{M_{\rm b}}{2 {\rm M_{\rm \odot}}} \Big)^{-2/3} \\
\Big( \frac{M_{\rm BH}}{4 \times 10^6 {\rm M_{\rm \odot}}} \Big)^{1/3}\Big( \frac{1- e_{\rm b}^2}{1 - e_{\rm BH}^2} \Big)^{1/2} {\rm pc},
\end{gathered}
\end{equation}
where e$_{\rm b}$ is the orbital eccentricity of the binary star.  This is due to general relativistic, or 
Schwarzschild apsidal, precision in the binary star \citep{holman97,blaes02,holman06}.

\section{The effect of mass loss on the photometric appearance of red giant branch stars} \label{ehbform}

In this section, we quantify the effects of collisions and, specifically, the subsequent tidal stripping on 
the photometric appearance of 
RGB stars.  Our primary concern is the stellar evolution, so we do not specify the nature of the impactor.  
Instead, we look at how discrete mass loss events affect the subsequent time evolution of the 
RGB luminosity and radius.  We note here that our calculated collision rates are sufficiently high that 
the underlying RGB populations could be significantly affected, and these stars dominate the light 
distribution in an old stellar population.  Due to the high extinction toward the Galactic centre, for 
example, these are the only (``old'') stars that can be observed.

Stars of initial mass\footnote{In this section, unless otherwise stated, mass, luminosity and radius are 
expressed in solar units, and our chosen unit of time is $10^6~$yr} $0.3 \lap M/M_{\odot} \lap 10$ become RGBs 
stars.  Exhaustion of hydrogen in the core is followed by burning of hydrogen to helium in a thin shell; the 
luminosity at the beginning of this evolutionary stage, i.e. at the base of the giant branch (BGB), is 
approximated by \citep[e.g.][]{1989ApJ...347..998E}:
\begin{equation}
\label{eqn:LBGB}
L_{\rm BGB} \approx \frac{2.15M^2+0.22M^5}{1+1.4\times10^{-2}M^2+5\times10^{-6}M^4},
\end{equation}
where M is the mass of the star.  The mass of the helium core gradually increases, and the star's 
radius and luminosity shoot up as the star climbs the giant branch.  
In stellar evolution models, the radius and luminosity during the giant phase are determined 
almost uniquely by the mass, $M_{\rm c}$, of the (helium) core.
Approximate relations for $0.17 \lap M_{\rm c}/M_{\odot} \lap 1.4$ are \citep{2006epbm.book.....E}:
\begin{eqnarray}
\label{eqn:LRRG}
L_{\rm GB} & \approx & \frac{10^{5.3}M_{\rm c}^6}{1 + 10^{0.4}M_{\rm c}^4 + 10^{0.5}M_{\rm c}^5}, \\
R _{\rm GB} & \approx & M^{-0.3}(L_{\rm GB}^{0.4}+0.383L_{\rm GB}^{0.76}).
\end{eqnarray}
The dominant energy source during these evolutionary phases
is the CNO cycle, and so the rate of energy production is tied to the rate of increase of the core mass by
\beq
\label{eqn:Lpp}
L_{\rm GB}  \approx 1.44 \times 10^{5}  \times \frac{dM_{\rm c}}{dt}.
\eeq
Given an initial core mass, Equations~\ref{eqn:LRRG}-\ref{eqn:Lpp} can be solved for the dependence of 
$M_{\rm c}$, $L_{\rm GB}$ and $R_{\rm GB}$ on time.
The surface temperature is given by the Stefan-Boltzmann law (however, at late times, the 
maximum radius is limited by mass loss).  
The initial core mass -- that is, the mass of the hydrogen-exhausted
helium core at the time the main sequence turnoff is reached -- is given by \citep{2006epbm.book.....E}:
\beq
M_{\rm c} \approx \frac{0.11 M^{1.2} + 0.0022M^2 + 9.6 \times 10^{-5} M^4}
{1 + 0.0018M^2 + 1.75\times 10^{-4} M^3}.
\eeq
For instance, for $M = (1,2,3)$ M$_{\odot}$, the initial core mass is
$M_{\rm c}/M_{\odot} = (0.11,0.26,0.43)$ and the initial core mass ratio is
$M_{\rm c}/M_{\odot} = (0.11,0.13,0.14)$.

For stars with masses $0.3 \lap M/M_{\odot} \lap 0.5$ the evolution on the RGB is interrupted when 
$M_{\rm c}$ has grown to approximately $M$.  Low mass stars, with $0.5 \lap M/M_{\odot} \lap 2$, 
develop degenerate He cores on the RGB and $3\alpha$  reactions initiate a violent nuclear
runaway, called the helium-flash.  Intermediate mass stars, with $2 \lesssim M/M_{\odot} \lesssim 10$,
develop non-degenerate He cores, which growth only slightly on the RGB before the He burning begins.  
The luminosity at the tip of the giant branch (TGB) is roughly $L_{\rm TGB} \approx L_{\rm BGB} + 2 \times 10^3$.

Mass loss can significantly alter the evolution of an RGB star.  For instance, it can 
prevent the star from igniting He, and therefore terminate the star's evolution at an 
earlier stage.  In order to study the response of RGB stars to episodes of intense mass loss, 
we used the stellar evolution code~STARS \citep{eggleton71,pols+95}.  
We compute evolution tracks for stars with various ZAMS masses, applying the \citet{dejager+88} 
prescription for the mass loss rate due to stellar winds.  
At some point during the RGB phase, we switch to a constant mass loss regime at the rate 
$10^{-5}\msun~\rm yr^{-1}$.  During this time we turn off changes in composition due to nuclear burning, 
while retaining energy production.  When the stellar mass is reduced to a certain value, $M_{\rm f}$, we
halt the rapid mass loss regime, and continue with "normal" stellar evolution by turning the 
de~Jager mass loss rate prescription back on, which in turn allows for composition changes to 
re-initiate. This numerical procedure allows us to study the evolution of stars undergoing 
an episode of mass loss on a time scale much shorter than their nuclear time scale, as is the case for 
an RGB star that is tidally stripped of its envelope.

\begin{figure}
\begin{center}
\includegraphics[width=0.73\linewidth,angle=0]{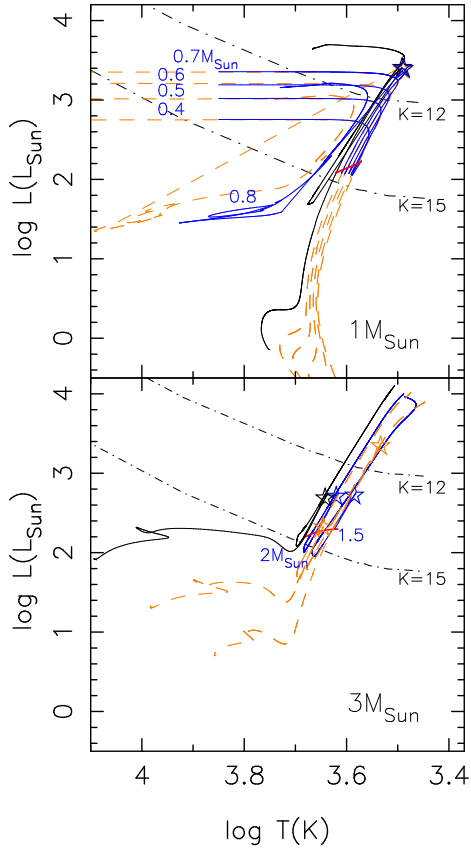}
\end{center}
\caption[Evolution tracks for stars undergoing an episode of mass loss on the giant branch]{Evolution tracks for stars with 
ZAMS masses of $M = 1 M_{\odot}$ and $3 M_{\odot}$ undergoing an 
episode of mass loss on the giant branch.  The total stellar mass, $M_{\rm f}$, remaining after some fraction of 
the RGB envelope was (instantaneously) expelled is shown in blue.  The star's evolution during the mass 
removal phase is shown by the red line, with the blue lines depicting the subsequent evolution.  
Stellar evolution tracks for unperturbed stars are shown via the black lines.  For comparison, for each model, 
we also show via the dashed orange lines the Eggleton stellar evolution tracks for unperturbed stars with 
$M = M_{\rm f}$.  The star symbols indicate the point on the H-R diagram where stars ignite He.  The 
dot-dashed lines are K=12, 15 contours obtained using giant colours and bolometric corrections from 
\citet{johnson66}.  
Stars with $K >15$ are too faint to be resolved at the MW Galactic centre.  For this plot, we assume 
a distance of $8~\rm kpc$ and an extinction of $A_{\rm K}=3$, suitable to the line of sight extending from the 
Sun to the Galactic Centre.
\label{fig:fig1}}
\end{figure}

Figure~\ref{fig:fig1} displays stellar evolution tracks on the RGB, after different amounts of envelope 
stripping.  Stars will not reach the TGB if the total stellar mass is smaller (due to stellar winds) than 
the He core mass at the TGB of the progenitor star.  In this case, the star will instead climb the giant branch 
until the mass of its hydrogen exhausted He core is essentially equal to its total stellar mass, at which point it 
will turn off the giant branch to the blue at the same luminosity as other stars of the same total mass.  The 
star will fail to ignite He and will subsequently cool off to become a white dwarf 
with an He core.  If the mass left in the star is instead large enough that its He core mass 
can reach that of the parent star at the moment of He ignition, then the remnant 
will also ignite He at roughly the same luminosity as its (unperturbed) parent star.  In this scenario, the star 
eventually cools off to become a white dwarf with a CO core. 
Figure~\ref{fig:fig1} shows that, for stars of total mass $0.8 \lesssim M/M_{\odot} \lesssim 1.5$, 
to significantly alter the evolution and make the TGB fainter requires $M_{\rm f} \lesssim 0.7 M_{\odot}$.
For $M \gtrsim 2 M_{\odot}$, we require instead $M_{\rm f} \approx 0.3 M_{\odot}$ to be removed or, 
equivalently, more than $90\%$ of the star's envelope \citep[e.g.][]{pradam09}.

We find that the overall stars' evolution is relatively insensitive to the luminosity at which the mass loss 
episode occurs.  Following mass removal, a star will nearly return to its initial luminosity on the RGB, 
but will move onto the point on the Hayashi track corresponding to its new mass.  This is independent 
of the luminosity at which tidal stripping occurs.  While the luminosity and evolutionary timescales 
remain virtually unchanged, the new radius of a stripped RGB star will increase (or decrease; see below) 
significantly as it reaches a new state of thermal equilibrium.

For $M_{\rm c}/M \gap(\lap) 0.2$, the star responds to mass loss by becoming smaller (larger).
In Figure~\ref{fig:fig4}, the dashed line corresponds to an He core mass equal to 0.2M, 
approximately the mass ratio below which RGB stars respond to tidal stripping by expanding.  Stars with 
masses $M \gtrsim 2 M_{\odot}$ never rise above this critical mass ratio and will always expand on
losing mass.  Lower mass stars, on the other hand, have $0.1 \lap M_{\rm c}/M \lap 0.15$ when they 
first leave the main sequence, and will expand upon losing mass.  Farther up the giant branch, 
the core mass increases relative to the total stellar mass, and these stars should shrink upon losing mass.

\begin{figure}
\begin{center}
\includegraphics[width=.8\linewidth,angle=-90.]{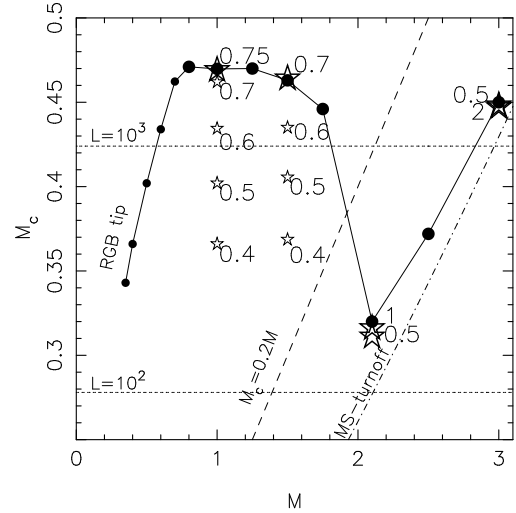}
\end{center}
\caption[The He core mass at the TGB as a function of the ZAMS mass.]{The He-core mass at the TGB is shown as a 
function of the ZAMS mass.  The filled circles show the 
results for unperturbed single stars.  The star symbols show the hydrogen-exhausted He core mass for stars 
undergoing an episode of rapid mass loss on the giant branch.  The symbols are larger for stars that ignite He.  The 
indicated masses show the final stellar mass that remains after mass removal.  The dashed line delineates the boundary 
between canonical stars (filled circles) that respond to mass loss by becoming bigger ($M_{\rm c}/M_{\odot} < 0.2$) or 
smaller ($M_{\rm c}/M_{\odot} > 0.2$).  The dot-dashed line shows the mass of the hydrogen-exhausted He core at the time 
corresponding to the main sequence turnoff.  All units are assumed to be solar.
\label{fig:fig2}}
\end{figure}

Figure~\ref{fig:fig2} shows the mass of the hydrogen exhausted core, $M_{\rm c}$, at the TGB as a function 
of the ZAMS mass for both unperturbed and stripped stars.  For unperturbed stars, there is a deep minimum 
around $M = 2 M_{\odot}$ which corresponds approximately to the transition between stars with 
degenerate, $M \lesssim 2 M_{\odot}$,  and non-degenerate, $M \gtrsim 2 M_{\odot}$, He cores at the beginning 
of the giant branch \citep{HS55}.  Comparing the He core mass at the TGB with that at the time corresponding to the 
main sequence turnoff (dot-dashed line in Figure~\ref{fig:fig2}), we see that for $M\gtrsim 2 M_{\odot}$ the majority 
of the He core mass needed for ignition of He-burning is already in place before the star reaches the giant branch.  
A star with $M \gtrsim 2M_{\odot}$ that undergoes a strong mass loss episode on the RGB will still experience a 
core-He burning phase and reach nearly the same luminosity as its parent star at the TGB.  Therefore, for 
$M \gtrsim 2 M_{\odot}$, mass loss will have little effect on the luminosity corresponding to the TGB, unless most of 
the envelope is suddenly removed near the BGB.  With that said, however, mass loss during the RGB phase 
could still make the tip of the asymptotic giant branch (AGB) non-negligibly fainter (for example, see the lower 
panels in Figure~\ref{fig:fig1}).

\section{Application to Local Group Galactic Nuclei} \label{app}

In this section, we review the available observations for the central nuclear regions of our four 
Local Group galaxies, namely the MW, M31, M32 and M33, in order to characterize their physical 
properties for input to the rate estimates presented in Section~\ref{rates}.  We further present the 
calculated collision rates for each nucleus as a function of distance from the cluster centre, using the 
observed NSC properties.

\begin{figure}
\begin{center}
\includegraphics[width=\columnwidth]{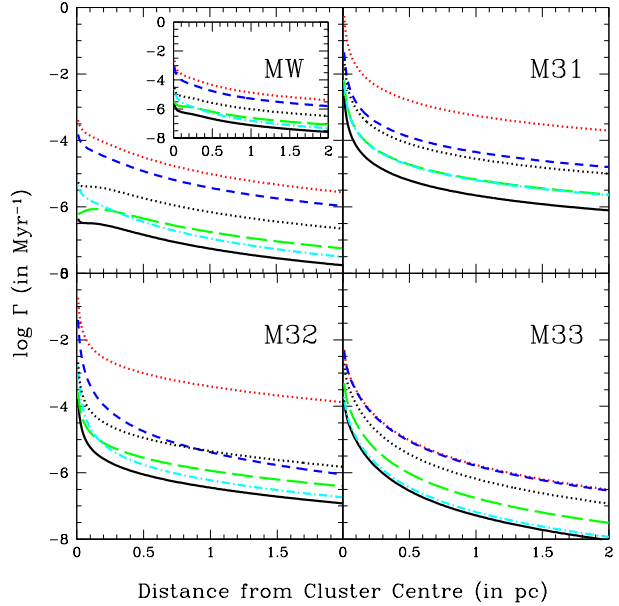}
\end{center}
\caption[Individual rates of MS+MS, MS+RGB, MS+WD, WD+RGB, BH+RGB and 1+2 collisions for the MW, M31, M32 and M33]{The MS+MS (black solid lines), 
MS+RGB (red dotted lines), MS+WD (green long-dashed lines), WD+RGB (cyan dot-dashed lines), BH+RGB (dotted black lines) and 1+2 (blue dashed 
lines) collision rates are shown as a function of distance from the cluster centre (or SMBH) for the MW (top left panel), M31 (top right panel), 
M32 (bottom left panel) and M33 (bottom right panel).   These are the rates for a \textit{particular} object to undergo a collision.  
For comparison, we also show our results for the MW 
assuming the density profile given by Equations 1 and 4 in \citet{merritt10}, via an additional 
inset in the top left panel.  The assumptions used as 
input to each rate equation in Section~\ref{rates} are discussed in Section~\ref{app}.  
\label{fig:fig3}}
\end{figure}

\begin{figure}
\begin{center}
\includegraphics[width=\columnwidth]{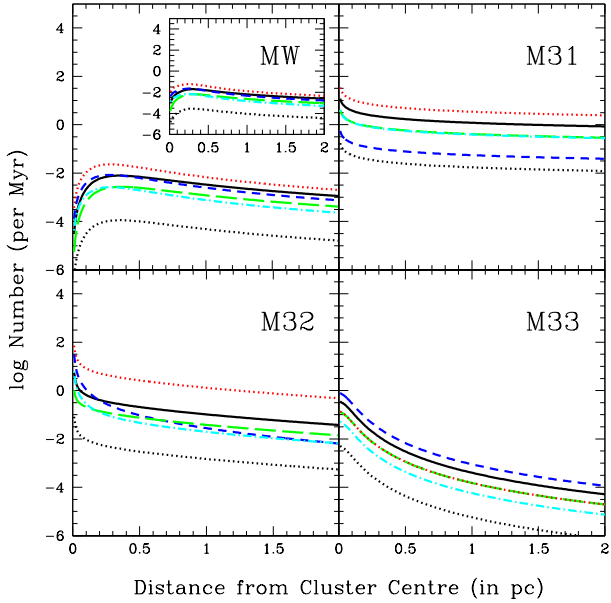}
\end{center}
\caption[Total Rates of MS+MS, MS+RGB, MS+WD, WD+RGB, BH+RGB and 1+2 collisions for the MW, M31, M32 and M33]{The total rates of 
MS+MS (black solid lines), MS+RGB (red dotted lines), MS+WD (green long-dashed lines), WD+RGB (cyan dot-dashed lines), 
BH+RGB (dotted black lines) and 1+2 (blue dashed lines) collisions are shown as a function of distance from the 
cluster centre (or SMBH) for the MW (top left panel), M31 (top right panel), M32 (bottom left panel) and M33 (bottom right panel).   These 
are the total rates for \textit{any} pair of objects to undergo a collision.  For 
comparison, we also show our results assuming the density profile given by Equations 1 and 4 in \citet{merritt10}, via an additional 
inset in the top left panel.  In making this figure, we adopt a 
bin size of 0.01 pc, such that the y-axis corresponds to the number of collision products in each 0.01 pc radial interval, over a 100 Myr 
time period.  The assumptions used as input to each collision rate are the same as in Figure~\ref{fig:fig3}.
\label{fig:fig4}}
\end{figure}

Our main results are shown in Figures~\ref{fig:fig3} and~\ref{fig:fig4}, which show the collision rates (both for a 
particular object and the total integrated rates) for the MW (top left insets), M31 (top right insets), M32 (bottom left insets) and 
M33 (bottom right insets), as a function 
of distance from the cluster centre or SMBH.  The solid black, dotted red, dashed blue, long-dashed green, dotted black and 
dot-dashed cyan lines show, respectively, the 
rates of MS+MS, MS+RGB, MS+WD, WD+RGB, BH+RGB and 1+2 collisions.  

Figure~\ref{fig:fig3} shows the individual rates for a 
\textit{specific} object to experience direct collisions with other objects of a given type, and must hence be multiplied by an 
appropriate fraction.  
 We multiply the rates of MS+MS, MS+RGB, MS+WD, WD+RGB, BH+RGB and 1+2 collisions by factors of (1-f$_{\rm b}$)f$_{\rm MS}$, 
(1-f$_{\rm b}$)f$_{\rm MS}$, (1-f$_{\rm b})$f$_{\rm MS}$, (1-f$_{\rm b})$f$_{\rm RGB}$, (1-f$_{\rm b})$f$_{\rm RGB}$ and 
(1-f$_{\rm b}$), respectively, where f$_{\rm b}$ is the fraction of objects that 
are binaries and f$_{\rm MS}$ is the fraction of single stars on the main-sequence.  We require 
1 $=$ f$_{\rm MS}$ + f$_{\rm RGB}$ + f$_{\rm WD}$ + f$_{\rm NS}$ + f$_{\rm BH}$, where f$_{\rm RGB}$, f$_{\rm WD}$, f$_{\rm NS}$ and 
f$_{\rm BH}$ are the fractions of single RGBs, WDs, NSs and BHs, respectively.  We set f$_{\rm b} =$ 0.01, f$_{\rm MS} =$ 0.89, 
f$_{\rm WD} =$ 0.10 and f$_{\rm RGB} =$ 0.01 \citep[e.g.][]{leigh07,leigh09,maeder09,leigh11b}.  These fractions are 
only approximate, but are representative of what we find upon integrating over a Kroupa initial mass function \citep{kroupa95,kroupa95b}, 
over the appropriate mass ranges for an old stellar population.  For the fractions of NSs and BHs, we
set f$_{\rm NS} =$ 0.01 and f$_{\rm BH} =$ 0.001 \citep[e.g.][]{alexander05,hopman06}.  

To estimate the binary fraction for 
each NSC, f$_{\rm b}$, we first calculate the expected average hard-soft boundary in the NSC for a typical binary.
We then calculate the fraction of the log-normal input period distribution, taken from \citet{raghavan10} for solar-type binaries
in the field, below this hard-soft boundary, $f_{\rm P}$.  We assume that, if the full period distribution could be occupied, this would
yield a binary fraction of 50\%.  The (hard) binary fraction is then given as the product f$_{\rm b} =$ 0.5f$_{\rm P}$.
The resulting binary fractions are lower in clusters with hard-soft boundaries at smaller orbital periods (i.e. clusters with larger
velocity dispersions), as is consistent with observed star clusters \citep[e.g.,][]{sollima08,milone12,leigh15}.  This gives
f$_{\rm b} =$ 0.010, 0.002, 0.020 and 0.060 for, respectively, the MW, M31, M32 and M33.  Finally, for the mean binary orbital
separation, we adopt the cluster hard-soft boundary.

Figure~\ref{fig:fig4} shows the total rates for \textit{any} pair of objects to undergo a collision.  These total rates are calculated 
by multiplying each collision rate in Figure~\ref{fig:fig3} by the number of objects (MS, RGB or WD) in a given radial bin.  We adopt a 
bin size of 0.01 pc, such that the y-axis corresponds to the number of collision products in each 0.01 pc 
radial interval, over a 100 Myr time interval.  We multiply each rate by the calculated 
number of objects of the given type within a given radial bin, found by multiplying the mean density in each 
bin by its volume.  An additional factor is included to correct for the fraction of objects of the desired type.  That is, 
we multiply the numbers of MS+MS, MS+RGB, MS+WD, WD+RGB, MS+NS, BH+RGB and 1+2 collisions by (1-f$_{\rm b}$)f$_{\rm MS}$, 
(1-f$_{\rm b})$f$_{\rm RGB}$, (1-f$_{\rm b})$f$_{\rm WD}$, (1-f$_{\rm b})$f$_{\rm WD}$, (1-f$_{\rm b})$f$_{\rm NS}$, 
(1-f$_{\rm b})$f$_{\rm BH}$ and f$_{\rm b}$, respectively.  We then calculate the total number of collision/merger products 
for each mechanism by integrating numerically over the 2-D and 3-D density profiles.  That is, we sum over all radial bins to 
estimate the total number of collision products expected in a 100 Myr time interval, from 0.1 pc (i.e., the current 
location of the inner blue stars) to 2 pc (i.e., a rough estimate for the relevant size of the 
surrounding old population) from the cluster centre.  The results are shown in Table~\ref{table:one} for 
all NSCs in our sample.

In making Figures~\ref{fig:fig3} and~\ref{fig:fig4}, we assume a mean single star 
mass and radius of 0.3 M$_{\odot}$ and 0.3 R$_{\odot}$, respectively, which are suitable for an old stellar 
population.\footnote{We do not integrate over a present-day stellar mass function in calculating the collision rates, due 
mainly to uncertainties in the cluster age and initial stellar mass function for the NSCs in our sample.  For example, 
some authors have argued for a top-heavy initial mass function in the Galactic Centre \citep[e.g.][]{paumard06}.}  
All RGB stars, WDs, NSs and BHs are assumed to have masses of, respectively, 1 M$_{\odot}$, 
1 M$_{\odot}$, 2 M$_{\odot}$ and 10 M$_{\odot}$.  All RGB stars are assumed to have radii
of 10 R$_{\odot}$, and all binaries have orbital separations equal to the hard-soft boundary.  Hence, gravitational focusing
is typically negligible for all collisions involving RGB stars (except if BHs are involved) and binaries, but is always significant for MS+MS, 
MS+WD, MS+NS and BH+RGB collisions.  We caution that the 
rates shown in Figures~\ref{fig:fig3} and~\ref{fig:fig4} are order-of-magnitude estimates, 
and are highly sensitive to our assumptions for the input parameters.  For example, an increase in the mean 
binary orbital separation by a factor of only 2 translates into an 
increase in both the corresponding rates and numbers by a factor of 8.  Thus, binary destruction via 1+2 
collisions could be (or have been in the past) much more efficient than Figures~\ref{fig:fig3} and~\ref{fig:fig4} would 
suggest, since we assume only very compact binaries.  


\begin{table*}
\caption{Total number of collision products expected in a 100 Myr time interval}
\begin{tabular}{|c|c|c|c|c|c|c|c|}
\hline
Galaxy          &      MS+MS      &      MS+RGB       &       MS+WD       &         WD+RGB    &      MS+NS       &        BH+RGB       &      1+2     \\
\hline
Milky Way    &     7.1 $\times$ 10$^1$     &    1.8 $\times$ 10$^2$    &      2.5    $\times$ 10$^1$    &   2.0 $\times$ 10$^1$    &   5.0 $\times$ 10$^0$    &   1.0 $\times$ 10$^0$     &    6.5 $\times$ 10$^1$       \\
M31              &     3.2 $\times$ 10$^4$     &     9.3 $\times$ 10$^4$     &     1.1 $\times$ 10$^4$     &     1.1 $\times$ 10$^4$      &       2.2 $\times$ 10$^3$    &   4.6 $\times$ 10$^2$      &   1.5 $\times$ 10$^3$     \\
M32              &     4.3 $\times$ 10$^3$    &     5.5 $\times$ 10$^4$       &       1.4 $\times$ 10$^3$       &     2.0 $\times$ 10$^3$      &         2.6 $\times$ 10$^2$    &   6.2 $\times$ 10$^1$  &    7.9 $\times$ 10$^3$        \\
M33              &     1.3 $\times$ 10$^4$      &       5.0 $\times$ 10$^3$     &     4.9 $\times$ 10$^3$     &        1.9 $\times$ 10$^3$      &       9.7 $\times$ 10$^2$    &   1.8 $\times$ 10$^2$    &   4.3 $\times$ 10$^4$    \\
\hline
\end{tabular}
\label{table:one}
\end{table*}


\subsection{The Milky Way} \label{MW}

At the heart of the Milky Way (MW), there resides an NSC of mass 
$\sim$ 9 $\times$ 10$^6$ M$_{\odot}$ ($<$ 100 arcsec) hosting a super-massive black hole 
of mass $\sim$ 4 $\times$ 10$^6$ M$_{\odot}$ \citep[e.g.][]{genzel96,chatzopoulos15}.
We assume a pressure-supported nucleus with an isothermal velocity dispersion 
$\sigma_{\rm 0} =$ 100 km/s \citep{merritt13}.  Using Equation~\ref{eqn:rinf}, this yields 
an influence radius for the central SMBH of $\sim$ 1.7 pc.  

The MW NSC is thought to have a core \citep[e.g.][]{merritt10}, hence we adopt the stellar density profile of \citet{stone15}:
\begin{equation}
\label{eqn:rho}
\rho(r) = \frac{\rho_{\rm 0}}{(1 + r^2/r_{\rm c}^2)(1 + r^2/r_{\rm h}^2)},
\end{equation}
where r is the distance from the origin, r$_{\rm c}$ is the core radius and r$_{\rm h}$ is the half-mass radius.  Equation~\ref{eqn:rho} 
has been chosen to produce a flat core inside a roughly isothermal cluster.\footnote{These assumptions are consistent 
with the observational constraints available for the MW NSC \citep[e.g.][]{merritt10}.  However, we note that, formally, 
a certain amount of anisotropy is needed for there to exist a core (in a Keplerian potential).}  Here, the core radius is given by:
\begin{equation}
\label{eqn:core}
r_{\rm c} = \frac{\sigma_{\rm 0}}{2{\pi}G\rho_{\rm 0}}.
\end{equation}

For comparison, we also show our results assuming the density profile given by Equations 1 and 4 in \citet{merritt10}, 
via an additional inset in Figures~\ref{fig:fig3} and~\ref{fig:fig4}.  As we will show, both of these assumptions for the 
density profile yield very comparable radial collision profiles.  The central density is assumed to be 
10$^6$ M$_{\odot}$ pc$^{-3}$, and we adopt a half-mass radius of 
r$_{\rm h} =$ 2.5 pc \citep{merritt13} along with a mass-to-light ratio of 2.  We use Equation~\ref{eqn:gamma10} 
to calculate all collision rates in the MW (with $\Sigma =$ 0), appropriate for a 
pressure-supported roughly isothermal nuclear environment, and transition smoothly between these formulae 
as gravitational-focusing becomes significant.  Note that we set M$_{\rm d} =$ 0, since there is no significant 
Keplerian disk component outside 0.1 pc.  The results are shown in Figures~\ref{fig:fig3} 
and~\ref{fig:fig4} by the black lines.

In the MW, the collision rates are sufficiently high that all types of collision products should be 
present, at least within $\lesssim$ 1 pc of the Galactic Centre.  This predicts 
non-negligible numbers of MS+MS collision products and hence blue stragglers, as 
illustrated in Figure~\ref{fig:fig4} and Table~\ref{table:one}.  More specifically, over a 1 Gyr period, 
on the order of 710 $\sim$ 10$^3$ MS+MS collisions should occur, which is roughly 0.1\% of the total 
stellar mass in this region.  
After an initial rise from r $=$ 0, the number of collision products remains roughly constant with 
increasing distance from the SMBH.  As shown in Figure~\ref{fig:fig3}, the 
predicted collision rates for a particular object show a near monotonic decrease with increasing distance 
from the cluster centre, after a relatively constant inner core that extends out to only $\sim$ 0.2 pc.
The density profile taken from \citet{merritt10}, while quantitatively very similar, shows a slightly steeper initial 
drop followed by a very similar monotonic profile.  The total collision rates for any pair of objects, however, show a 
rapid increase out to $\sim$ 0.3 pc, 
followed by a monotonic decrease.  This is the case for both of our assumed density profiles, which yield 
very similar collision rate profiles, as shown in Figure~\ref{fig:fig4}.  Thus, we might expect a relatively 
weak central concentration of blue stars in the MW, assuming a collisional origin (ignoring any possible 
mass segregation of collision products).  

The core radius given in Equation 4 of \citet{merritt10} is much smaller 
than the SMBH influence radius, such that the stellar orbits within this core should be roughly Keplerian.  
This is what creates a central drop in some collision numbers within the core, as shown in Figure~\ref{fig:fig4}.  
However, this inner core is such a small 
fraction of the total volume over which we integrate the collision rates that this correction should have a 
relatively negligible impact on the total numbers of collision products.  

Very roughly, the numbers of 1+2 collisions that occur over a $\gtrsim$ 10 Gyr period 
should be comparable to the total number of binaries in the MW NSC, for our assumed binary fraction and 
considering only hard binaries.  The hard-soft boundary in the MW NSC corresponds to the binary 
components being in (or nearly in) contact, due to the very high velocity dispersion.  Hence, the probability of 
a direct collision or merger occurring during 
an encounter involving such a hard binary is nearly unity for small impact parameters \citep[e.g.][]{leigh12}.  The 
collision product should expand post-collision by a factor of $\sim$ a few \citep[e.g.][]{sills97,fregeau04}, driving 
the remaining binary components to merge when their radii overlap at periastron.  Conversely, encounters 
involving soft binaries should have a high probability of being dissociative, or of significantly increasing the 
binary separation post-encounter such that the time-scale for a subsequent encounter (with an even higher 
probability of dissociation) becomes very short.  
Thus, only (near) contact binaries should survive in the Galactic Centre for any significant amount of time although 
many will also be destroyed, and our estimate for the binary fraction in the MW NSC 
(i.e., f$_{\rm b} =$ 0.01) could be an over-estimate of the true binary fraction.  

With that said, it seems unlikely that the disk of O- and B-type stars observed at $\sim$ 0.1 pc from 
Sgr A* has a collisional origin.  First, although collisions should be dissipative, there is no obvious reason why the 
collision products should 
arrange themselves into a disk-like configuration in a spherical pressure-supported nucleus.  Second, given 
the average stellar mass in the 
MW NSC, multiple collisions would be required to form these massive O- and B-type stars, which must 
have occurred over a very short interval of time, since O- and B-stars have very short lifetimes.  Figure~\ref{fig:fig3} 
suggests that the rate of collisions involving individual MS stars is too low for this to be the case.  

Interestingly, however, the rate of MS+WD collisions is sufficiently high that a non-negligible supply of gas could be 
supplied to the nucleus via this mechanism (especially assuming the density profile in \citet{merritt10}).  This is 
because collisions between stars and WDs should act to ablate 
the star, nearly independent of the cluster velocity dispersion (see Section~\ref{discussion} for more details regarding 
the underlying physics responsible for this mechanism for gas liberation; \citep{shara77,shara78,regev87}).  This mechanism 
could supply on the order of $\lesssim$ 10$^3$ M$_{\rm \odot}$ in gas every $\sim$ 1 Gyr in the MW NSC.  This is because, using 
the density profile of \citet{merritt10}, on the 
order of $\sim$ 100 MS/RGB+WD collisions happen every 100 Myr (see Table~\ref{table:one}).  Assuming that most MS stars 
undergoing MS+WD collisions are close to the turn-off, which is of order $\sim$ 1 M$_{\odot}$ for an old population, then 
Figure~\ref{fig:fig2} (see the dot-dashed line) suggests that the He core mass should comprise a small fraction of the total 
stellar mass ($\lesssim$ 0.2 M$_{\odot}$ or 20\%) when the collision occurs.  Hence, for our purposes, a reasonable 
assumption for the mean mass in gas liberated per MS/RGB+WD collision is $\sim$ 0.8 M$_{\odot}$ (i.e., the envelope, or all 
material outside of the degenerate core).  Assuming 10$^3$ MS/RGB+WD collisions per 1 Gyr, this predicts 
$\sim$ 800 M$_{\odot}$ $\sim$ 10$^3$ M$_{\odot}$ in gas should be supplied to the inner nucleus every Gyr due to 
MS+WD collisions alone.  

The above back-of-the-envelope calculation suggests that MS+WD collisions alone cannot supply 
enough gas to form stars.  This is because the time required to accumulate enough gas to form an 
actual disk is much longer than the gas cooling time (of order a few to a few hundred 
years, for either optically thick or thin disks), which must be less than roughly 3 times the local dynamical 
time-scale \citep{nayakshin06,levin07,chang07}.  Thus, if star formation were to occur only from gas supplied 
by MS+WD collisions, then it should only occur after nearly a Hubble Time.  Given that stellar 
mass-loss in the surrounding old population should likely supply gas to the inner nucleus at a comparable 
or even higher rate 
\citep[e.g.][]{chang07}, and the fact that the gas must not be accreted by the central SMBH before it 
fragments to form stars, we conclude that, while MS+WD collisions could contribute non-negligibly to the 
gas supply in the inner nucleus, this mechanism seems by itself insufficient to form stars and an 
additional source of gas is needed.

As shown in Section~\ref{ehbform}, nearly the entire RGB envelope must be stripped to significantly affect the 
photometric appearance of an RGB star.  Given that more than one collision is typically required to strip an RGB of its 
envelope \citep[e.g.][]{bailey99,dale09,amaroseoane14}, the 
derived rate for a given RGB star to encounter other MS stars is too low for collisions to contribute significantly 
to the observed paucity of giants (as shown by the dotted red line in Figure~\ref{fig:fig3}).  Assuming instead an 
isothermal density profile that starts to diverge near the SMBH, 
only in the inner $\lesssim$ 0.1 pc could the rate become sufficiently high for any given RGB star to encounter more 
than a single MS star within $\lesssim$ 1 Gyr (i.e. the typical lifetime of an RGB star).  
Although we expect 
BH+RGB collisions to be more effective at stripping the RGB envelope on a per collision basis \citep{dale09}, 
we find that the rate of BH+RGB collisions is likely too low for more than a handful of RGB stars to have been 
fully stripped.  We conclude that it seems difficult to account for any missing RGB stars in the Galactic Centre 
via collisions.     

This is consistent with the general picture described in \citet{merritt10}, which suggests that the stellar density 
of the underlying (not yet observed) old population traces that of the observed RGB stars, and has always been 
low.  In the absence of star formation, a low initial density remains low for a long time, since the relaxation time is 
very long.  Figures 4 and 5 in \citet{merritt10} show that a model with a core can reproduce the observed number 
counts of RGB stars in the MW.  Our fiducial model for the MW also has a core, which is partly responsible for the 
low predicted numbers of collisions involving RGB stars.  Thus, the low observed density of RGB stars could be 
consistent with being a natural consequence of a low (initial) density in the underlying old population.  This picture is 
consistent with the low rate of RGB collisions found here, which would not be able to account for any inferred 
paucity of RGB stars.

To summarize, the predicted rates of single-single collisions are too low to have significantly affected the 
photometric appearance of the MW NSC.  This is the case for stars belonging to any evolutionary stage.  In particular, 
MS+MS collisions should have produced $\lesssim$ 1\% of the total stellar mass in collision products after a Hubble 
time.  Collisions involving individual RGB stars occur too infrequently for the RGB envelope to have been affected 
significantly, and any paucity of RGB stars to be observed.  The rate of single-binary collisions, on the other hand, is 
sufficiently high that very few binaries should remain at the present day, if any.  Finally, the rate of MS+WD collisions 
is insufficient to supply enough gas to form stars and, while this mechanism for gas accumulation could contribute 
non-negligibly to a single burst of star formation, an additional source of gas is also needed.

\subsection{M31} \label{m31}

The M31 nucleus is a distinct stellar system at the centre of its host late-type spiral galaxy, with a surface brightness profile that 
rises significantly above the background potential at r $<$ 10 pc.  The nucleus shows an asymmetric double-lobed 
structure within r $<$ 3 pc of the central SMBH, which is consistent with the diffuse eccentric disk model 
of \citet{tremaine95}.  Here, the nucleus consists of an old stellar population, but at even smaller radii (r $<$ 0.6 pc; 
relative to the photocentre of the surrounding bulge component) there exists an UV-bright cluster of blue stars whose
origins are unknown.  Unlike the younger and hotter O- and B-stars observed in our Galactic Centre, these 
appear to be A-type stars \citep{bender05,lauer12}.  Below, we elaborate further on the details of the M31 nucleus.

At the centre of M31, there lurks an SMBH of mass $\sim$ 1.4 $\times$ 10$^8$ M$_{\odot}$ 
\citep{dressler84,dressler88,kormendy88,richstone90,kormendy99,bender05}.  
The central blue cluster, called P3, is surrounded by two over-densities of stars, called P1 and P2 
\citep{lauer93}, which reside 
on either side of P3 with a separation of $\sim$ 1.8 pc \citep{bender05}.  These 
two ``lobes'' are distinct from P3 both in terms of their stellar content and 
kinematics \citep[e.g.][]{lauer12}.  P1 and P2 are much redder in 
colour, while P3 contains a significant ultraviolet excess 
\citep[e.g.][]{nieto86,king92,king95,lauer98,brown98}.  A velocity dispersion of $\sim$ 250 km/s 
has been measured in P2 along the same line of sight to P3, with a maximum 
of 373 $\pm$ 48 km/s on the anti-P1 side of the central blue cluster \citep{bender05}.  
Using Equation~\ref{eqn:rinf}, the mean velocity dispersion gives an influence 
radius for the central SMBH of $\sim$ 9.6 pc.  P3, on the other hand, has the highest 
velocity dispersion measured to date, with 
a central dispersion of 1183 $\pm$ 200 km/s \citep{bender05}.

The most likely explanation for the double-lobed structure of P1+P2 is 
an eccentric disk viewed in projection.  This was originally proposed by 
\citet{tremaine95}, who argued that P1 and P2 are unlikely to be two distinct 
star clusters caught during the final stages of a merger, since the merger would 
occur within $\lesssim$ 10$^8$ years by dynamical friction \citep{lauer93,emsellem97}.  
To reconcile this, 
\citet{tremaine95} proposed that both nuclei or ``lobes'' are part of the same eccentric 
disk of stars.  The observations can be explained if the brighter lobe, P1, is located 
at a farther distance from the central SMBH and is the result of stars lingering near 
apocentre \citep[e.g.][]{statler99,kormendy99,bacon01}.  Conversely, the fainter lobe, P2, 
can be accounted for if it corresponds 
approximately to pericentre and the disk density increases toward the central SMBH.  Hence, 
the SMBH dominates the central potential, such that stars in the surrounding disk 
should follow approximately Keplerian orbits.\footnote{If the disk has a mass $\gtrsim$ 10\% that of the SMBH, 
then self-gravity is required to keep the disk aligned \citep{statler99}.}    
\citet{peiris03} later refined the eccentric disk model, taking advantage of more recent 
ground-based spectroscopy to help constrain their improved model.  The models were 
used to predict the kinematics that should be observed via the Ca triplet in HST spectra 
of P1+P2, and the model predictions are in excellent agreement with the 
data \citep{bender05,lauer12}.

Our collision rate estimates in Section~\ref{collisions} are based on alterations to the 
formulation provided in \citet{goldreich04}, originally derived to treat collisions within 
protoplanetary disks.  This approach can be almost directly applied in M31, 
since the total mass of the Tremaine disk is M$_{\rm d}$ $\sim$ 3 $\times$ 10$^7$ M$_{\odot}$ 
\citep{bender05} (assuming all 1.0 M$_{\odot}$ stars and a mass-to-light ratio of 
5.7, which is appropriate for a bulge population \citep{tremaine95}).  This is nearly 
an order of magnitude smaller than the mass of the central SMBH.  Hence, here we 
can ignore any self-gravity within the disk \citep[e.g.][]{statler99,emsellem07}.

We use Equation~\ref{eqn:gamma10} to calculate the collision rates in M31 with $\rho =$ 0, 
appropriate to a Keplerian nuclear disk (as in the MW, we transition smoothly between 
these formulae as gravitational-focusing becomes significant).  The disk has a scale length and 
height of, respectively a $=$ 1.8 pc and h $=$ 0.7a.\footnote{The M31 nuclear disk is observed 
to satisfy the relation 1 - h/a $>$ 0.3 \citep{lauer93}.}  We assume a constant surface mass 
density $\Sigma =$ M$_{\rm d}$/($\pi$a$^2$), where M$_{\rm d} =$ 3 $\times$ 10$^7$ M$_{\odot}$ 
\citep{bender05}.  This is because, as shown in Figure 17 of \citet{lauer98}, the central surface 
brightness profile in M31 cannot be described by a simple analytic function. 
Critically, this neglects the torus-like structure of the M31 nucleus, and over-estimates the surface 
mass density at small radii close to the SMBH (ignoring the inner blue disk).  However, as we will show, 
the collision rates are sufficiently high in this region that collisions would have transformed the inner 
nucleus over the last few Gyrs.  Thus, the absence of a significant background of old stars in the inner 
nucleus at present is not necessarily indicative of an absence in the past.  Consequently, we assume 
that the presently-observed torus-like structure was once a diffuse disk extending further in toward the 
central SMBH, and include any collision products calculated to have occurred there in our estimates.  However, 
we note that if this assumption is incorrect, and the presently observed torus was always present, then 
there is a true inversion in the stellar density at small radii and hence the collision rates here would be 
negligible, dropping to zero at small r instead of continually rising all the way down to r $=$ 0, as shown 
in Figure~\ref{fig:fig3}.

As in the MW, the collision rates in M31 are sufficiently high that all six types of collision products 
should be present in non-negligible numbers over the entire extent of the disk (i.e. $\gtrsim$ 2 pc), as 
illustrated in Figure~\ref{fig:fig4}.  The radial dependence of the collision rate is weak beyond $\gtrsim$ 1pc, 
but becomes more significant in the inner nucleus.  
Specifically, the predicted collision rates at $\sim$ 0.1 pc should outweigh those at $\sim$ 1 pc by $\lesssim$ an 
order of magnitude, for all six types of collision products, as shown in Figure~\ref{fig:fig4}.  

Only in the inner $\lesssim$ 0.1 pc is the rate of MS+RGB collisions ever sufficiently high for individual RGB stars 
to have undergone enough collisions ($\sim$ 100) over 100 Myr to unbind most of the RGB envelope, and 
hence to cause an observed paucity of RGB stars.  This is the case for MS+RGB collisions alone, however, 
since multiple collisions involving the same main sequence star and other MS stars are less common.  
This suggests that any BSs present in M31 are likely only slightly more massive than the MS turn-off, 
and hence should appear as A-type stars, as observed for the inner blue disk \citep{lauer12} (and not 
O- and B-type stars, unlike in M33 and M32; see below).

The numbers of 1+2 collisions over a $\gtrsim$ 100 Myr interval should 
exceed the total numbers of binaries in M31, for our assumed binary fraction of f$_{\rm b} =$ 0.01 and 
considering only hard binaries.  The velocity dispersion in M31 is so high that the hard-soft boundary 
corresponds to the binary components being in contact.  Hence, for small impact parameters, the 
probability of a direct collision or merger occurring during any direct 1+2 encounter involving a hard binary 
is approximately unity \citep[e.g.][]{leigh12}.  Conversely, the probability of dissociation during encounters 
involving soft binaries is very high.  Thus, even (most) contact binaries should not have survived until the 
present-day, and our estimate for the binary fraction in the M31 nucleus could be an over-estimate of the true 
binary fraction.  

MS+WD collisions should liberate on the order of 
10$^5$ M$_{\rm \odot}$ in gas every $\sim$ 1 Gyr, which can subsequently be used to form 
young (blue) stars.  This is because on the 
order of $\sim$ 1.1 $\times$ 10$^4$ MS+WD collisions happen every 100 Myr (see Table~\ref{table:one}).  
Hence, following the same assumptions as in the preceding section for the MW, if 
1.1 $\times$ 10$^5$ MS+WD collisions occur per 1 Gyr, this predicts 
$\sim$ 8.8 $\times$ 10$^4$ M$_{\odot}$ $\sim$ 10$^5$ M$_{\odot}$ in gas should be supplied to the 
inner nucleus every Gyr due to MS+WD collisions alone.  In M31, however, the critical mass needed for 
fragmentation is 10$^4 <$ M$_{\rm crit}$/M$_{\odot} <$ 10$^5$ for a distance from the 
cluster centre 0.1 $\le$ r/pc $\le$ 3 (see Figure 7 in \citet{chang07}).  As in the MW, the 
gas accumulation times needed to satisfy this critical mass requirement are much longer 
than the gas cooling time.  Thus, these results suggest that star formation should occur 
on the order of every $\sim$ 1 Gyr in M31, if the only supply of gas to the inner nucleus is 
MS+WD collisions.  Once again, the rate of mass loss due to stellar evolution is, very roughly, 
comparable to the rate at which gas is supplied by MS+WD collisions.  Hence, the actual time-scale 
for disk fragmentation could be much shorter than this simple calculation suggests, and 
gas from stellar evolution in the surrounding old population should contribute non-negligibly to the 
total gas mass available for fragmentation.

We note that, in M31, the central velocity dispersion rises toward r $=$ 0 and can even exceed 
1000 km/s interior to $\lesssim$ 0.1 pc.  This should decrease the rates of MS+MS, MS+WD, MS+NS 
and BH+RGB collisions relative 
to what is shown in Figures~\ref{fig:fig3} and~\ref{fig:fig4}, by reducing the significance of gravitational 
focusing.  Additionally, given 
such high relative velocities at impact, some direct MS+MS collisions could 
completely unbind or ablate the collision product, leaving behind a puffed-up cloud of gas and 
dust in its place \citep{spitzer66,spitzer67}.  Similarly, grazing or off-centre collisions could 
(eventually) leave behind a very rapidly rotating collision product.  In any event, the collisional velocities 
at impact are sufficiently high in M31 very close to the central SMBH that gas can be supplied to the 
NSC not only via MS+WD and WD+RGB collisions, but also via 
MS+MS and MS+RGB collisions (in addition to mass loss due to stellar evolution).  Additionally, 
we caution that, if a non-negligible fraction of MS+MS collisions ablate the product completely, then 
we might not expect any central rise in the numbers of blue stragglers, but rather a dip in the numbers.  
Alternatively, if the products of MS+MS collisions are not ablated, then the deposited kinetic 
energy at impact will serve to puff up the collision product, reducing the time-scale for subsequent 
collisions to occur.  Immediately after a collision, the product shoots up the Hayashi track before 
contracting back down to the main-sequence \citep[e.g.][]{sills97,sills01}.  Hence, at least temporarily, 
the collision product 
will be brighter and redder than a normal MS star of the same mass.  Interestingly, the photometric 
appearance of the object observed at the nominal centre of P3, called S11, is qualitatively consistent 
with this general picture, since it is the brightest and reddest object in the central blue cluster \citep{lauer12}.

To summarize, the predicted rate of MS+MS collisions is sufficiently high in the inner $\lesssim$ parsec 
of the M31 nucleus that most, if not all, of the mass of the inner blue disk in M31 could be composed of MS+MS 
collision products, or blue stragglers.  Importantly, these collisions would have dissipated orbital angular 
momentum and, given the disk-like nature of the M31 NSC, the thin disk-like structure of the inner blue excess 
is in general consistent with a collision origin.  Note that the additional concentration of collision products within 
the plane of the surrounding disk is unique to the M31 nucleus, relative to the other NSCs in our sample.  
We caution that MS+MS collisions could become ablative in the inner 
$\lesssim$ 0.1 pc due to the very high velocity dispersion here, which predicts impact velocities with enough 
kinetic energy to exceed the binding energy of any collision product.  This would contribute to the total gas supply 
in the inner M31 nucleus, but likely not enough to form the inner blue disk in a burst of star formation.  On the other 
hand, MS+WD collisions, which should be largely ablative nearly independent of the impact velocity, could 
contribute significantly to the total gas reservoir in the inner nucleus.  Together with stellar evolution-induced 
mass loss in the surrounding old population, this could supply enough gas to form stars in much less than a 
Gyr.  Finally, the rate at which individual RGB stars collide with other objects is sufficiently high in the inner 
$\lesssim$ 0.1 pc that enough collisions could have occurred to completely unbind the RGB 
envelope, causing an observed paucity of RGB stars in the inner M31 nucleus.

With the above results in mind, we note that \citet{yu03} found from a similar analysis of analytic 
collision rates calculated for the M31 nucleus that collisions alone cannot account for the blue excess observed 
at r $<$ 0.6 pc.  We note, however, that \citet{yu03} took the (currently) observed present-day density of the old 
population at face value, and did not account for the possibility that the Tremaine disk once extended inward 
to smaller radii.  What's more, \citet{yu03} did not have the more recent results of \citet{bender05} and \citet{lauer12}, 
which provide, respectively, better constraints for the properties of the blue cluster P3 and better spatial 
resolution for the entire inner M31 nucleus.  Overall, our estimates for the collision rates in M31 are in decent 
agreement with those calculated by \citet{yu03}, but the additional information provided by more recent 
observational studies has allowed us to perform a slightly more focused analysis based on a more up-to-date 
understanding of the structure of the inner nucleus.  This accounts for the different conclusions reached by 
\citet{yu03} for the M31 nucleus, relative to those presented in this paper.

\subsection{M32} \label{M32}

The nearby elliptical galaxy M32 is home to the smallest of the three SMBHs known to exist 
in the Local Group, with a mass $\lesssim$ 2.5 $\times$ 10$^6$ M$_{\odot}$ 
\citep{tonry84,verolme02,vandenbosch10}.  The properties and even detection of this SMBH are, 
however, controversial \citep{merritt13}.  Surrounding this putative central SMBH is a rotating 
disk of stars with a density $>$ 10$^7$ M$_{\odot}$ pc$^{-3}$ at r $<$ 0.1 pc 
\citep{walker62,lauer98}.  Beyond this, an additional stellar component is 
present, with an effective radius of $\sim$ 6 pc and a total mass 
$\sim$ 3 $\times$ 10$^7$ M$_{\odot}$ \citep{kormendy99,graham09}.  This represents 
$\sim$ 10\% of the total galaxy mass, such that the NSC in M32 contains a much larger 
fraction of the total galaxy luminosity than is typical for early-type galaxies 
\citep[e.g.][]{cote06}.  The stellar populations characteristic of the central NSC are younger 
than is normal for the rest of the galaxy, with a mean age of $\sim$ 4 Gyr in the central 
nuclear region and rising to $\sim$ 8 Gyr at larger radii \citep{worthey04,rose05,coelho09}.  
Even though the nucleus is a two-component system, with a dominant or primary disk component 
embedded in a more spherically-distributed pressure-supported secondary component, there is 
no observed break in the properties of these stellar populations as a function of radius \citep{seth10}.  

The typical velocity dispersion in the M32 nucleus is $\sim$ 60 km/s, but rises to $\sim$ 120 km/s 
very close to the central SMBH \citep{seth10}.  Using Equation~\ref{eqn:rinf}, this yields an 
influence radius for the central SMBH of only $\lesssim$ 0.7 pc.  Thus, we use Equation~\ref{eqn:gamma10} 
with $\Sigma =$ 0 to calculate the collision rates in M32, appropriate to a dynamically hot pressure-supported 
nuclear cluster, and transition smoothly between these formulae as gravitational focusing becomes important.  
As in the MW, however, we correct for the influence of the central SMBH by replacing the isothermal velocity dispersion 
with the local velocity dispersion given by Equation~\ref{eqn:sigloc}, with $\sigma_{\rm 0} =$ 60 km/s and 
M$_{\rm BH} =$ 2.5 $\times$ 10$^6$ M$_{\odot}$.  

For the surface mass density profile, we adopt the best-fitting solution to 
the V-band surface brightness profile from \citet{lauer98}, which is a Nuker-law fit of the form:
\begin{equation}
\label{eqn:sbm32}
\Sigma(r) = 2^{(\beta-\gamma)/\alpha}{\Upsilon}\Sigma_{\rm 0}\Big( \frac{r_{\rm b}}{r} \Big)^{\gamma}\Big[1 + \Big( \frac{r}{r_{\rm b}} \Big)^{\alpha}\Big]^{(\gamma-\beta)/\alpha},
\end{equation} 
where $\Upsilon =$ 2.0 is the V-band mass-to-light ratio (in M$_{\odot}$/L$_{\odot}$), $\alpha =$ 1.39, 
$\beta =$ 1.47, $\gamma =$ 0.46, r$_{\rm b} =$ 0".47 and $\Sigma_{\rm 0}$ is the central surface mass 
density in L$_{\odot}$pc$^{-2}$, calculated from the V-band surface brightness $\mu_{\rm 0} =$ 12.91 
given in \citet{lauer98} assuming a distance to M32 of 770 kpc.  In calculating the collision rates using 
Equation~\ref{eqn:gamma10}, we assume $\Sigma$ $=$ 0 (since there is no Keplerian disk component).  Assuming 
spherical symmetry, we can obtain 
the stellar density distribution from the observed surface brightness profile using Equation 3.65a in \citet{merritt13}:
\begin{equation}
\label{eqn:rhofromsb}
\rho(r) = -\frac{1}{\pi} \int_r^{\infty} \frac{d\Sigma}{dR} \frac{dR}{\sqrt{R^2 - r^2}}.
\end{equation}
which is independent of any assumptions for the gravitational potential.  Equation~\ref{eqn:rhofromsb} is the 
classical Abel inversion of a projected surface brightness profile \citep[e.g.][]{bracewell65}.  Note that, in M32, the break 
radius corresponds almost exactly to the outer radial limit we adopt for the calculated collision rates, 
or r$_{\rm b} \sim$ 2 pc.  Inside the break radius, the approximation $\rho$(r) $\sim$ $\Sigma$(r)/r is 
reasonable for power-law galaxies, where 
$\Sigma$(r) is the observed surface brightness profile, as given by Equation~\ref{eqn:sbm32}.  


As shown in Figure~\ref{fig:fig3}, the predicted collision rates for a particular object show a rise toward 
r $=$ 0, increasing by $\lesssim$ two orders of magnitude over the inner $\sim$ 1 pc.  For WD+RGB and 1+2 collisions, 
however, the central increase is much less significant.  Upon integrating 
numerically, the predicted numbers of collision products remain approximately constant with 
increasing distance from the galaxy centre, as shown in Figure~\ref{fig:fig4}.  We see only a slight rise in 
the numbers of collision products at very small radii $\lesssim$ 0.1 pc, similar to what is predicted for M31.  Thus, 
we do not expect a significant central concentration of blue stars in M32 (albeit perhaps a mild one at 
$\lesssim$ 0.1 pc that would be difficult to identify observationally, at least currently), assuming a collisional 
origin (ignoring any possible mass segregation of collision products).  Interestingly, M32 is the only 
galaxy in our sample for which this is the case, while also being the only galaxy in our sample 
that does not show any observational evidence for a central blue excess.  

Beyond $\gtrsim$ 0.1 pc, the integrated numbers of collision products are high 
relative to the other galaxies in our sample (ignoring M31); the predicted 
numbers of MS+MS, MS+WD and MS+RGB collisions produced over a Hubble time should be of order a few percent 
of the total number of stars 
in the M32 nucleus, as shown in Table~\ref{table:one}.  This naively suggests that the observed light distribution in 
the M32 NSC, and hence its stellar mass function, could have been non-negligibly affected by collisions.  However, The 
MS+MS collision rates for a given MS star are too low for any massive 
O- and B-type stars to have formed from multiple collisions.  And yet, the collision rates are high enough that collisions 
could be connected with the lack of recent star formation in M32.  Collisions could perhaps act to inhibit star 
formation by, for example, colliding with protostars sufficiently early on in their formation that 
the protostellar embryos are destroyed.


As in both the MW and M31, the rate of 1+2 collisions in the M32 
nucleus is sufficiently high that most binaries should have been destroyed by the present-day (since the 
hard-soft boundary is at only 0.04 AU), 
and a total mass in gas of order $\sim$ 10$^3$ M$_{\odot}$ should be supplied to the inner nucleus 
every $\sim$ 100 Myr.  This is because on the 
order of $\sim$ 1.4 $\times$ 10$^3$ MS+WD collisions happen every 100 Myr (see Table~\ref{table:one}).  
Hence, following the same assumptions as in the preceding sections, if 
1.4 $\times$ 10$^4$ MS+WD collisions occur per 1 Gyr, this predicts 
$\sim$ 0.7 $\times$ 10$^4$ M$_{\odot}$ $\sim$ 10$^4$ M$_{\odot}$ in gas should be supplied to the 
inner nucleus every Gyr due to MS+WD collisions alone.  This is enough gas for star formation to occur, 
ignoring any other potentially complicating effects.  More generally, it predicts that non-negligible amounts of 
gas should be present in the nucleus at any given time, roughly independent of any recent star 
formation or the details of gas heating/cooling.


Additionally, as shown in Figure~\ref{fig:fig3}, the rates of MS+RGB collisions are 
only sufficiently high in the inner r $\lesssim$ 0.1 pc for any given RGB star to undergo multiple (up 
to $\sim$ 100 or more) collisions within 100 Myr.  This suggests that only in the very inner nucleus could RGB 
stars in M32 be fully stripped, and appear as EHB stars.  This predicts a paucity of giants in the very inner 
$\lesssim$ 0.1 pc of the M32 nucleus.

In summary, the single-single collision rates are sufficiently high that the numbers of collision products 
produced over a Hubble time should be of order a few percent of the total number of stars 
in the M32 nucleus.  These rates are roughly independent of distance from the cluster 
centre, apart from a brief but sharp rise near the origin.  Hence, only in the inner $\lesssim$ 0.1 pc of the nucleus 
could the collision rates ever be sufficiently high to create a (significant) blue excess via MS+MS collisions.  
Individual RGB stars could have undergone multiple collisions only within r $<$ 0.1 pc, with 
sufficient frequency for their envelopes to become fully stripped and their photometric appearance significantly 
affected.  Finally, the rates of MS+WD collisions are sufficiently high to supply 
significant ($\lesssim$ 10$^4$ M$_{\odot}$ every 1 Gyr) gas to the nucleus.  This predicts non-negligible quantities of gas to 
be present in the M32 nucleus at present (assuming they did not already form stars; but see Section~\ref{sf}).

Finally, we note that \citet{yu03} obtained similar results to what we report here, namely that collisions 
alone are unlikely to produce any colour gradients in M32, consistent with observations.  However, we point out that the 
collision rates are sufficiently high for a significant fraction of stars in the cluster to have undergone at least one collision over a 
Hubble Time.  This could significantly impact the stellar mass function and hence luminosity profile.  The weak 
dependence of the collision rate on clustercentric distance suggests that this effect would 
affect the overall colour of the M32 nucleus roughly uniformly (with only a slight central blue excess).  This 
could directly affect any age determination assigned to the M32 nucleus from stellar population synthesis models 
(likely causing an under-estimation).  Only in 
the inner $\lesssim$ 0.1 pc could the collision rates ever become sufficiently high for individual objects to undergo 
multiple collisions.  Here alone, our results predict a paucity of RGB stars.  Thus, similar to \citet{yu03}, we 
conclude that collisions are unlikely to be able to cause the appearance of an age gradient in M32.  

\subsection{M33} \label{m33}

M33 is a nearby spiral galaxy lacking a significant bulge component.  Intriguingly, this could be connected 
with the lack of any detected SMBH in M33.  \citet{gebhardt01} placed an upper limit of 1500 M$_{\odot}$ on the 
mass of any SMBH that could reside in the nucleus, using three-integral dynamical models fit to Hubble Space 
Telescope WFPC2 photometry and Space Telescope Imaging Spectrograph spectroscopy.  
The NSC at the centre of M33 is 
extremely compact, with a central density approaching that of M32 \citep{kormendy93,lauer98}.  The 
central velocity dispersion is only 21 km/s, and reaches central mass densities of at least 
2 $\times$ 10$^6$ M$_{\odot}$ pc$^{-3}$ \citep{lauer98}.  Assuming an SMBH mass of 1500 M$_{\odot}$, 
this yields an influence radius of only 0.01 pc, using Equation~\ref{eqn:rinf}.   Importantly, this is only an 
upper limit for the SMBH mass.  There could be no SMBH in M33.  Indeed, the presence of an additional strong 
X-ray source \citep{long96,dubus99} 
in the nucleus so close to the putative SMBH suggests against the presence of a massive 
central BH, since SMBHs are very effective at tidally disrupting X-ray binaries and even ejecting their compact remnant 
companions from the NSC and even host galaxy \citep{leigh14,giersz15}.

We use Equation~\ref{eqn:gamma10} with $\Sigma =$ 0 to calculate the collision rates in M33, which are 
appropriate to a dynamically hot pressure-supported nuclear cluster.  We note that in M33 we find 
v$_{\rm esc} >$ $\sigma$ for all types of collisions, and gravitational focusing must always be taken into 
account.  For the surface mass density profile, we again adopt the best-fitting solution to 
the V-band surface brightness profile from \citet{lauer98}, which includes an analytic core in this case:
\begin{equation}
\label{eqn:sbm33}
\Sigma(r) = {\Upsilon}\Sigma_{\rm 0}\Big[1 + \Big( \frac{r}{a} \Big)^{2}\Big]^{-0.745},
\end{equation} 
where $\Upsilon =$ 0.4 is the V-band mass-to-light ratio (in M$_{\odot}$/L$_{\odot}$), a $=$ 0.1 pc is the 
approximate half-power radius, $\Sigma_{\rm 0}$ is the central surface mass density in L$_{\odot}$pc$^{-2}$, 
calculated from the V-band surface brightness $\mu_{\rm 0} =$ 10.87 given in \citet{lauer98} assuming a 
distance to M33 of 785 kpc.  Although Equation~\ref{eqn:sbm33} is consistent with the observed surface 
brightness profile in M33, we note that \citet{lauer98} also considered a model with an inner cusp, and 
\citet{carson15} adopted Sersic models.

As in the preceding section, when calculating the collision rates using 
Equation~\ref{eqn:gamma10}, we assume $\Sigma$ $=$ 0 (since there is no Keplerian disk component).  We adopt 
the stellar density distribution provided in Equation 2 of \citet{merritt01b}.  This yields results 
that are nearly identical to what we find upon integrating over the observed surface brightness profile of the nucleus.  
That is, assuming spherical symmetry, we can obtain 
the stellar density distribution from the observed surface brightness profile using Equation~\ref{eqn:rhofromsb}, as 
before, and requiring that the luminosity density at r $=$ 1 pc is 10$^6$ L$_{\odot}$ pc$^{-3}$.  Note that, in M33, 
the core radius corresponds almost exactly to the inner radial limit we adopt for the 
calculated collision rates, or a $\sim$ 0.1 pc.  Outside the core radius, the surface brightness profile approximately 
resembles a power-law form, and the approximation $\rho$(r) $\sim$ $\Sigma$(r)/r is reasonable.  


Figure~\ref{fig:fig3} illustrates that the predicted collision rates for a particular object decrease rapidly with 
increasing distance from the galaxy 
centre, dropping by about three orders of magnitude over the inner $\sim$ 1 pc.  In Figure~\ref{fig:fig4}, 
we see a similar behaviour and the total rates for any pair of objects follow a monotonic decrease with increasing 
distance from the galaxy centre, dropping by about two orders of magnitude over the inner $\sim$ 1 pc.  As shown 
in Table~\ref{table:one}, the predicted total 
numbers of MS+MS, MS+WD and MS+RGB collisions over a Hubble Time is greater than the total number of 
stars in the central (i.e., the core) NSC.  The number of 1+2 collisions are sufficiently high that almost if not every binary 
should have undergone 
a direct encounter over the cluster lifetime.  However, in general, a non-negligible number of binaries should survive 
in M33, given its low velocity dispersion of $\sim$ 21 km/s, which yields a hard-soft boundary of 0.6 AU.  This is 
very close to the hard-soft boundaries of typical Milky Way globular clusters \citep{harris96}, which are known to 
have binary fractions on the order of a few percent \citep[e.g.][]{sollima07,sollima08,milone12}.  

A total mass in gas comparable to the total NSC mass in the core should be supplied 
to the inner nucleus every $\sim$ 1 Gyr due to ablating MS+WD collisions.  This is because on the 
order of $\sim$ 4.9 $\times$ 10$^3$ MS+WD collisions happen every 100 Myr (see Table~\ref{table:one}).  
Hence, following the same assumptions as in the preceding sections, if 
4.9 $\times$ 10$^4$ MS+WD collisions occur per 1 Gyr, this predicts 
$\sim$ 2.5 $\times$ 10$^4$ M$_{\odot}$ $\sim$ 10$^4$ M$_{\odot}$ in gas should be supplied to the 
inner nucleus every Gyr due to MS+WD collisions alone.  This is less than the total mass of the M33 nucleus by about 
an order of magnitude.  This also implies that stars should have formed recently if the gas is able to become sufficiently 
dense to cool.  As in M32, this further predicts that significant gas 
should be present in the nucleus at any given time, roughly independent of any recent star 
formation or the details of gas heating/cooling.  
Finally, the collision rates are never high enough in M33 for individual RGB stars 
to undergo multiple MS+RGB collisions within 100 Myr.  Thus, we do not expect a paucity of RGB stars in 
the M33 nucleus.  


To summarize, the photometric appearance of the M33 nucleus should have also been significantly 
affected by collisions; the total number of collision products should be of order $\lesssim$ 10\% of the total 
number of stars in the inner nucleus.  An excess (by roughly an order of magnitude) of blue stars from MS+MS 
collisions is expected at r $\lesssim$ 0.1 pc.  
However, individual MS and RGB stars should not have undergone multiple collisions with any regularity.  Since 
of order $\sim$ 100 collisions are 
needed to fully strip the RGB envelope and significantly affect its photometric appearance, our results do not predict any 
observed paucity of RGB stars.  Unlike the other nuclei in 
our sample, however, binary destruction due to single-binary encounters should be comparably inefficient, due to 
the much lower velocity dispersion.  This predicts a binary fraction of order $\lesssim$ a few percent, and possibly a 
large number of X-ray binaries due to exchange interactions involving hard binaries and stellar remnants.

%
%

\section{Discussion} \label{discussion}

In this section, we discuss the implications of our results for the origins of the centrally concentrated 
blue stars observed (or not) in the four Local Group galactic nuclei considered here and, more generally, for 
observing exotic or enigmatic stellar populations in 
galactic nuclei, including blue straggler stars, extreme horizontal branch stars, young or recently formed 
stars, X-ray binaries, etc.  Our results suggest that collisional processes could 
contribute significantly to the observed blue excesses in M31 and M33, via several different channels.  Below, 
we highlight the relevant physics for the development of improved theoretical models, which are needed for 
comparison to existing and future observational data.

\subsection{Young stars and/or recent star formation} \label{sf}

The presence of very blue stars in the inner nuclear regions is typically argued to be evidence for a recent 
(in situ) burst of star formation.  This is because the timescale for such a cluster of young stars to inspiral into 
the nucleus via dynamical friction tends to exceed the age of the stars, for all but extremely close initial 
distances from r $=$ 0.  Many authors have proposed channels via which such in situ star formation might 
occur so close to a central SMBH \citep[e.g.][]{chang07,nayakshin06}.

Given the current state of the observations, we can neither rule out nor confirm an in situ star formation origin for 
any blue excesses observed in our sample, at least when considering individual nuclei.  Upon consideration 
of the entire sample at once, however, more stringent constraints can perhaps be placed.  For example, 
if the star formation occurs in a single burst, some fine-tuning must be required to produce 3 out of 4 galactic 
nuclei with recent star formation 
within $\lesssim$ 0.1 pc of the galactic centre at the current epoch.  This is because, relative 
to the lifetimes of the older populations in these NSCs, the young blue stars would live for only a 
very short time.  Thus, it is unlikely that all three nuclei were caught in the act of recent star formation, unless 
the process is continuous or episodic with a short recurrence time.  If correct, then significant star formation 
must have previously occurred in 
the nucleus as well.  It follows that the most recent burst would have to occur in the presence of a significant 
population of remnants orbiting within $\lesssim$ 0.1 pc.  This predicts that a high 
mass-to-light ratio could also be present along with any inner blue excess, and that significant X-rays could 
be observable due to gas accretion onto these inner remnants (if gas is still being supplied to the inner nucleus; 
see below).

Interestingly, as shown in Section~\ref{app}, the rates of MS+WD collisions are sufficiently high 
that a significant and steady supply of gas could be supplied to the inner nuclei.  This is because, unlike 
MS+MS collisions, collisions between MS stars and WDs do not require very high relative velocities to ablate the star 
\citep{shara77,shara78}.  The WD is sufficiently compact that, upon penetrating the star, a shock wave is sent 
through it, increasing the temperature in the outer layers by roughly an order of magnitude.  This in turn triggers 
CNO burning, which liberates enough energy to unbind most of the envelope.  The collision of a WD with a massive 
MS star is hence disruptive, with the energy source unbinding the MS star being due only in small part to the kinetic 
energy of the non degenerate star \citep{regev87}.  Thus, MS+WD collisions can supply 
gas to the nucleus independent of its velocity dispersion, which sets the typical relative velocity at infinity for direct 
collisions.  In the MW and M31, enough gas could be supplied to the inner NSC via MS/RGB+WD collisions to account 
for the mass observed in blue stars via star formation every $\lesssim$ 1 Gyr.  In the inner $\lesssim$ 0.1 pc of M31 
and M32, the rates of MS+WD 
collisions are the highest, such that the recurrence time for star formation in this scenario should be considerably 
shorter.   For comparison, Equation 8 in \citet{generozov15} gives the mass input from stellar winds, parameterized 
by the fraction $\eta$ of the stellar density being recycled into gas.  With a lower limit of $\eta \ge$ 0.02 (and taking into 
account the star formation efficiency), our results 
suggest that the rate of gas supplied by ablative WD collisions could be comparable to the rate of gas supplied by stellar 
winds in the surrounding old population.  This could represent a significant correction to steady-state models of 
gas inflow/outflow in galactic nuclei \citep[e.g.][]{quataert04,shcherbakov10,generozov15}.

We emphasize that MS+MS collisions should contribute little to the total gas supply in the nuclei in our sample.  This is 
because the velocity dispersions are sufficiently low that the total kinetic energy at impact is only ever a small fraction of 
the binding energy of the collision product.  This is the case even upon consideration of the entire spectrum of possible 
relative velocities at impact, appropriate for a Maxwellian velocity distribution.  Only in the inner $\lesssim$ 0.1 pc 
of the M31 nucleus does the velocity dispersion ever become sufficiently high for MS+MS collisions to ablate the collision 
product and contribute to the total gas reservoir.

If young stars are indeed formed from violently liberated gas during collisions then, in a sample of NSCs with 
inner blue stars, we would expect a correlation between the (integrated) NSC collision rate and the total 
mass in blue/young stars.  With this in mind, we can already say that M32 would be a significant outlier in such a 
relation, ignoring any over-looked effects that could potentially inhibit star formation preferentially 
in the M32 nucleus (e.g. the M32 NSC is by far the densest in our sample, such that collisions could even 
serve to inhibit star formation).  Additionally, 
this scenario predicts approximately solar abundance for such collisionally-derived 
young stars, without any significant helium enrichment (assuming that the collisionally-destroyed MS stars 
are members of the original old stellar population).  This is because only a small fraction of the total mass 
actually undergoes enhanced CNO burning during the collision.

\citet{generozov15} recently calculated steady-state one-dimensional hydrodynamic profiles for hot gas in 
the immediate vicinity of an SMBH.  The authors provide an analytic estimate for the "stagnation radius" in their 
Equation 14, defined as the distance from the SMBH at which the radial velocity of the gas passes through zero:  
\begin{equation}
\label{eqn:rstag}
r_{\rm s} = \frac{GM_{\rm BH}}{\nu(v_{\rm w}^2 + \sigma_{\rm 0}^2)}\Big( 4\frac{M(r < r_{\rm s})}{M_{\rm BH}} + \frac{13 + 8\Gamma_{\rm s}}{4 + 2\Gamma_{\rm s}} - \frac{3\nu}{2 + \Gamma_{\rm s}} \Big),
\end{equation}
where M(r $<$ r$_{\rm s}$) is the total stellar mass inside the stagnation radius, $\Gamma_{\rm s}$ is the two-dimensional power-law 
slope of the stellar light distribution (not to be confused with the collision rate), $\nu$ is the three-dimensional power-law 
slope of the gas inflow profile (defined by Equation 13 in \citet{generozov15}) and v$_{\rm w}$ is an effective heating 
parameter that describes the energy deposited from stellar winds, supernovae and BH feedback.  For an old 
stellar population, we expect v$_{\rm w} \sim$ 100 km/s.  Inside the stagnation radius 
gas flows inward, whereas outside this radius gas flows outward.  In the MW and M31, the velocity dispersion is sufficiently 
high that the stagnation radius lies outside the SMBH sphere of influence, and beyond our outer limit of integration when 
calculating the integrated number of collision products (i.e. 2 pc).  Thus, all of the gas liberated from MS/RGB+WD collisions 
in our calculations should quasispherically flow inward, until it circularizes at its angular momentum barrier.   In M32, 
we expect r$_{\rm s} \gtrsim$ r$_{\rm inf}$, and the numerical solution of the 1D hydro equations shows that r$_{\rm s} =$ 2 pc 
for v$_{\rm w} =$ 100 km/s (Aleksey Generozov, private communication).  
In M33, assuming that no SMBH is present, the stagnation radius will be close to zero since v$_{\rm w} \gg$ $\sigma_{\rm 0}$, as 
can be seen from taking the limit M$_{\rm BH} \rightarrow$ 0 in Equation~\ref{eqn:rstag}.  Numerical experiments confirm that r$_{\rm s} \ll$ 1 pc for v$_{\rm w} \sim$ 100 km/s (Aleksey Generozov, private communication).  Hence, any (hot) gas liberated by collisions should be lost to a quasispherical outflow.  Finally, 
we note that the preceding estimates for the stagnation radius assume that stellar winds dominate the energy injection rate.  However, 
in the inner nuclei of M31 and M32, the energy injection due to collisions could dominate, causing an increase in v$_{\rm w}$ and a subsequent decrease in the stagnation radius. 


In all nuclei in our sample, our results suggest that the 1+2 collision rate is sufficiently high that most 
stellar binaries should be driven to merge, destroyed (via collisions) or dissociated very early on in the cluster lifetime, compared to 
the age of an old stellar population ($>$ a few Gyr).  It follows that most binaries \textit{currently} present could be 
very young (mostly likely formed due to recent star formation).  This predicts that, if 
young binaries are observed in a galactic nucleus, this could be a smoking gun for recent star formation, since 
no other mechanism considered in this paper or presented in the literature should yield a bright blue 
star with a binary companion.  Star formation, on the other hand, is expected to yield very high 
binary fractions for massive stars \citep[e.g.][]{sana12}.  Provided that star formation occurred sufficiently 
recently, most of these binaries should still be present.  Observationally, unresolved binaries should broaden
the color distribution (at a given magnitude) along the inferred main-sequence, relative to a coeval population 
of single stars.  With that said, our results presented in Section~\ref{bmkl} suggest that the time-scales for 
Kozai-Lidov oscillations to operate are sufficiently short very close to a central SMBH that all binaries 
with initially high inclinations should merge on relatively short time-scales, leaving both binaries and merged 
binaries in young nuclear environments.  

To summarize, our results suggest that collisions could offer an additional pathway toward re-fueling an inner 
nucleus with gas for recent star formation.  MS+WD collisions could contribute significantly to the total gas supply needed 
to form the centrally concentrated blue stars in every nuclei in our sample, with the possible exception of M32 and 
perhaps M33 (due to a relatively small stagnation radius for gas inflow/outflow).  In nearly all cases, the relative velocities are 
not high enough to ablate the stars during MS+MS collisions, except within the inner $\sim$ 0.1 pc of the M31 nucleus where 
the velocity dispersion reaches $\sim$ 1000km/s.  Here, however, MS+MS collisions could also offer an additional source of 
gas for star formation, as originally suggested by \citet{spitzer66} and \citet{spitzer67}. During MS+WD collisions, on the other 
hand, the relative velocity at impact plays only a 
minor role in deciding how much gas is liberated.  This is because the strong surface gravity of the WD is primarily responsible 
for generating the high temperature shock front that triggers explosive CNO burning in the MS star, as opposed to the kinetic 
energies of the objects at impact (see \citet{shara77} and \citet{shara78}).  

\subsection{Missing giants and extreme horizontal branch stars} \label{ehb}

If the envelope of an RGB star is stripped, the hot core is revealed and the star settles onto the extreme horizontal 
branch in the cluster colour-magnitude diagram before dimming to join the white dwarf cooling sequence 
\citep[e.g.][]{davies98,amaroseoane14}.  If only a single grazing collision occurs, only a small fraction of the 
RGB envelope should end up unbound.  Although the photometric appearance of the resulting RGB star can still be 
slightly affected by a single mass loss episode, our results presented in Section~\ref{ehbform} suggest that almost all 
of the RGB envelope must be stripped before its subsequent evolution and photometric appearance are significantly 
affected, which should require multiple collisions.  
%

It is not clear how many MS+RGB collisions are needed for an RGB star to be collisionally stripped by MS stars to 
the point of forming an hot EHB star.  The number could range from on the order of 10 to hundreds \citep[e.g.][]{dale09}.  
Thus, the relevant rate here is that for a \textit{specific} RGB star to undergo collisions with MS stars (i.e. Figure~\ref{fig:fig3}), 
which is lower than the rate for \textit{any} RGB star to undergo collisions by a factor N$_{\rm RGB}$, 
where N$_{\rm RGB}$ is the total number of RGB stars.  The rates 
of MS+RGB collisions are sufficiently high in only the inner $\lesssim$ 0.1 pc of  M31 and M32 
that this mechanism could potentially produce EHB stars (and hence a paucity of RGB stars) in the inner nuclear 
regions.  Given that the 
EHB lifetime is only on the order of 10$^8$ years \citep[e.g.][]{maeder09,leigh11b}, however, 
it seems unlikely that any observed blue excess is due to EHB stars, since the implied rate of EHB formation 
must be unrealistically high.  Although BHs are expected to 
be able to remove a larger fraction of the envelope on a per encounter basis \citep{dale09}, Figure~\ref{fig:fig3} 
suggests that only in perhaps M33 could the rate of BH+RGB collisions be of comparable significance 
to MS+RGB collisions for removing RGB envelopes.  A 
deficit of late-type giants can perhaps be looked for observationally using the presence or absence of CO bandhead 
absorption to distinguish late-type from early-type stars (see \citet{genzel96} for more details).

\subsection{Blue straggler stars} \label{bss}

Blue straggler stars are rejuvenated main-sequence stars, identifiable in the cluster colour-magnitude
diagram as being brighter and bluer than the main-sequence turn-off \citep{sandage53}.  BSs are thought 
to form from a number of channels in both open and globular clusters, including direct collisions 
between main-sequence stars \citep[e.g.][]{shara97,sills97,sills01,leigh07,leigh11}, mass-transfer in 
binary systems from an evolved donor onto a normal MS star 
\citep[e.g.][]{mccrea64,mathieu09,geller11,geller13,knigge09,leigh11b} and mergers of MS-MS binaries 
\citep[e.g.][]{chen09,perets09}.  In M31, M32 and M33, the rate of MS+MS collisions is sufficiently high to supply, 
in as little as a few Gyrs or less, a total mass in MS+MS collision products that is a significant fraction of the total 
mass of the inner nuclear regions, and could thus contribute non-negligibly to the observed light distribution.  

With the exception of M33, the relaxation times of the NSCs in our sample tend to be on the order of a few 
Gyr or more \citep{merritt13}.  This is longer than the expected age of a typical collision product formed from an old 
population \citep[e.g.][]{sills97,sills01}.  It follows that mass segregation of collision products formed at larger radii 
into the inner nuclear regions should have only a minor impact on deciding the observed numbers of collision 
products here.

With that said, any tension due to an over-abundance of predicted BSs can perhaps be reconciled by the 
fact that the products of direct MS+MS collisions should be puffed up post-collision due to the deposition of 
kinetic energy \citep[e.g.][]{sills97,fregeau04}.  This increases the collisional cross-section and reduces the 
time-scale for subsequent collisions, which could act to strip the collision product of its inflated envelope.  This 
would serve to reduce the acquired mass in collision products.   The envelope should contract on a 
Kelvin-Helmholtz time-scale, which can be comparable to the time for subsequent collisions to occur.  Finally, 
we note that more massive collision products should have shorter lifetimes, which reduces their probability of 
actually being observed.  

If BSs are produced abundantly, then we might naively also expect an over-abundance of RGB stars, since 
the heavier BSs should evolve onto the giant branch on shorter time-scales than the rest of the MS 
population.  As explained in the next section, these evolved BSs might be rapidly destroyed via MS+RGB 
and WD+RGB collisions, however we do not expect this to be the case in the MW and M33.  And yet, in the 
MW NSC, Table~\ref{table:one} suggests that, over a 1 Gyr period, we expect on the order of 10$^3$ MS+MS 
collisions in the inner $\sim$ 2 pc.  Although this is only a small fraction of the total number of RGB stars, this excess 
is nonetheless puzzling in the MW NSC, since a paucity of giants has been observed, not an excess 
(but see the last paragraph in Section~\ref{MW}).  

Very roughly, the rates of 1+2 collisions are sufficiently high in every NSC in our sample that most primordial binaries 
should have been destroyed by the present-day (with the exception of M33).  It follows that our assumed binary fractions 
could be over-estimates, 
in spite of correcting for the fraction of hard binaries using the observed velocity dispersion.  More importantly, this 
suggests that not enough binaries should be present 
to contribute significantly to BS production via either binary coalescence or binary mass-transfer.  Only those 
binaries born in a near contact state have sufficiently small cross-sections (i.e. near contact) to survive for 
$\gtrsim$ 1 Gyr.  This is because the orbital separation corresponding to 
the hard-soft boundary is near contact in nuclei with such high velocity dispersions.  For 
very close binaries near contact, the probability of a merger or direct collision occurring during a 1+2 
interaction approaches unity \citep[e.g.][]{fregeau04,leigh12}.  Hence, most direct 1+2 encounters that do not 
result in a direct stellar collision should be dissociative, or at least widen the binary and reduce the time-scale for a 
subsequent (likely dissociative) encounter to occur.  It follows that the rate of direct 1+2 encounters 
should be approximately equal to the rate of binary destruction very early on in the cluster lifetime.  For those 
few near contact binaries that are able to survive for the bulk of the cluster lifetime, if stable mass-transfer is 
initiated then mass, energy and angular momentum conservation dictate that the binary orbit should 
(eventually) widen, significantly increasing the geometric cross-section and reducing the time-scale for a 
direct 1+2 encounter \citep[e.g.][]{leigh16}.  Similarly, if unstable mass-transfer is initiated, then a common 
envelope could form, 
which would also (at least temporarily) increase the geometric cross-section for collision and reduce the 
1+2 collision time.  Thus, we expect that the very few BSs that might form from binary mass-transfer should 
quickly lose their binary companions post-formation, due to dissociative or destructive 1+2 encounters \citep{leigh16}.  
The simple calculations performed here should be verified and better quantified in future studies, using more 
realistic binary fractions and distributions of binary orbital parameters.


Finally, we consider one additional mechanism for BS formation, which has not yet been considered.  This is 
the accretion of significant amounts of gas from the interstellar medium (ISM) onto a normal old MS star.  The gas 
could be supplied to the ISM by MS+WD and even MS+MS collisions (if the velocity dispersion is very high, which 
is the case in only M31 and only in its inner $\lesssim$ 0.1 pc; see Section~\ref{sf}), and/or 
mass loss from evolving stars in the surrounding old population.  The gas collects at the 
tidal truncation radius (if an SMBH is present) due to the focusing of gas particle orbits in an axisymmetric potential 
\citep{chang07}.  This gas is then channeled inward on a viscous time-scale, where we are assuming it is accreted by 
old main-sequence stars already orbiting there.  
As discussed by \citet{nayakshin06} for a similar scenario, as the gas density builds up, feedback from stars can serve 
to stop fragmentation 
of the gas, while accretion onto pre-existing old stars could proceed at very high rates.  Although this scenario 
requires an initial old and centrally-concentrated stellar population to provide the seeds for accretion, it is 
appealing since it ultimately requires a much smaller gas reservoir than the recent burst of star 
formation scenario, while also avoiding the special conditions required to induce the fragmentation of 
a giant molecular cloud in such an extreme environment so close to an SMBH.  We caution that more work needs 
to be done to quantify the expected accretion rates in this scenario, which are poorly known.

Thus, accretion onto pre-existing old MS stars also predicts 
an inner disk of blue stars in the M31, M32 and M33 nuclei, assuming that the accretion rates of the ISM onto the 
old stars are sufficiently high.  This mechanism is also appealing in that, unlike a recent burst of star formation, 
it is a \textit{continuous} mechanism for rejuvenating old MS stars.  Indeed, we might naively expect the mean BS mass 
(and hence luminosity) for BSs formed from this mechanism in the inner nucleus to correlate with the age of the 
surrounding old population, since older stars will have had longer to expel their winds into the ISM, and to accrete 
the polluted material.  This mechanism also predicts 
low binary fractions, a very low central mass-to-light ratio and (possible) surface abundance spreads throughout the 
cluster \citep[e.g.][]{lauer98,seth10}, depending on the quantity and composition of the accreted material.

We conclude that all nuclei in our sample should harbour an abundance of blue stragglers, formed mostly from MS+MS 
collisions.  Accretion onto old MS stars already orbiting in the inner nuclear regions offers an additional possible pathway 
toward BS formation.  Our results suggest that the rates of MS+WD collisions are sufficiently high to contribute significantly 
to the gas reservoir 
required for this mechanism to explain the observed blue stars.  These "non-binary mass transfer BSs" should have 
properties tightly correlated with those of the underlying old stellar population.

\subsection{Implications for inferred age spreads} \label{inferred}

As illustrated in Table~\ref{table:one}, our results predict non-negligible numbers of collision products formed 
over only a 100 Myr interval that can be a significant fraction of the total number of stars in the nuclear cluster.  To help 
distinguish blue stars formed via collisions from young stars and test for a true age gradient in a given NSC, 
particular attention can be put into deriving the radial dependence of the collision rate as a function of the host 
cluster environment.  The collision number 
profiles can be converted into corresponding surface brightness profiles, and from there into radial colour profiles.  
These collisional colour profiles can then be subtracted from the observed colour profiles, after multiplying the collisional 
colour profiles by an appropriate constant to ensure that no over- or under-subtraction occurs -- the normalization 
should be applied at large radii, or where the predicted collision rates are at their lowest.  If any 
residual colour gradient is present, this suggests that its origin is a recent burst of star 
formation.  If no residual colour gradient remains and the U-V or U-B colours turn out to be approximately 
constant as a function of distance from r $=$ 0, then the observed colour gradient is indeed consistent with 
a collisional origin.  This test can be performed on the Local Group nuclei when future observations with 
superior resolution capabilities become available and, in particular, more accurate velocity information. 

A number of collisional mechanisms considered in this paper could contribute to the appearance of a colour gradient, 
and should be accounted for when looking for a true age gradient.  
In every nucleus in our sample, our results predict non-negligible numbers of MS+MS collision products, or blue stragglers.  
In the inner $\lesssim$ 0.1 pc of M31 and M32, the collision rates are sufficiently high that individual 
RGB stars could undergo multiple 
collisions within 100 Myr.  If this were to result in a paucity of RGB stars and a corresponding excess of single EHB stars, then 
the contribution of RGB-stripping to observed colour gradients is two-fold; the loss of RGB stars reduces the total amount of 
red light and the production of EHB stars can increase the total amount of blue light.  Thus, the stripping of RGB 
stars could also contribute to the appearance of an age gradient, provided the numbers of MS+RGB, WD+RGB and 
BH+RGB collisions are extremely high and centrally concentrated.  

Finally, we note that future studies considering age gradients in nearby galactic nuclei should include the 
Local Group dwarf spheroidal galaxy NGC 205, which is known to harbor bright blue (and possibly young) stars.  We 
have not done so in this paper, since the rate of collisions is expected to be much lower than in the other NSCs in our 
sample \citep{valluri05}, due to a lower central surface brightness. 

\subsection{Binary mergers due to Kozai-Lidov oscillations} \label{bmkl}

The key point to take away from Equation~\ref{eqn:taukl} is that, for any binary within a distance r$_{\rm SC}$ 
from the centre of the galaxy with a sufficiently large angle of inclination relative to its orbital plane about the SMBH 
(i.e. i $\gtrsim$ 40$^{\circ}$), the 
Kozai-Lidov time-scale is much shorter than the cluster age, for any old (i.e. $\gtrsim$ a few Gyrs) stellar 
population.  For example, for a binary with a separation of 1 AU and component masses 
of 1 M$_{\rm \odot}$ on a circular orbit about the SMBH with a semi-major axis equal to the 
scale radius of the nucleus, the Kozai-Lidov time-scale given by Equation~\ref{eqn:taukl} is 4.8 $\times$ 10$^4$, 
1.0 $\times$ 10$^2$ and 1.5 $\times$ 10$^5$ years in, respectively, the MW, M31 and M32.  We also calculate 
the critical distance r$_{\rm SC}$ from the SMBH beyond which Kozai-Lidov oscillations are suppressed by general 
relativistic precession.  This is 0.04 pc, 0.15 pc and 0.04 pc for, respectively, the MW, M31 and M32.

These Kozai-Lidov time-scales are significantly shorter than the ages of the surrounding old stellar populations, 
as well as the collision times 
presented in Section~\ref{app}.  It follows that, over sufficiently long time-scales (much shorter than the cluster age), the 
rate of binary mergers due to Kozai-Lidov oscillations 
should be determined by the time-scale for single stars to scatter the binary orbital plane (and reduce the distance from 
the SMBH to $\lesssim$ r$_{\rm SC}$) into 
the active Kozai-Lidov domain.  Vector resonant relaxation acting to re-orient the binary orbital plane could contribute 
to either an increase or decrease in this rate (see \citet{antonini12b} for more details).  Regardless, in all nuclei in our 
sample, the rate of direct 1+2 collisions is 
sufficiently high relative to the age of the old stellar population that only very compact binaries should 
survive to the present-day, since the hard-soft boundary is near contact in these galactic nuclei due to their high velocity 
dispersions and the collision probability for a direct encounter with small impact parameter is close to unity for such compact 
binaries \citep{leigh12}.  The 
collision product should expand after the encounter by a factor of a few \citep[e.g.][]{sills97,fregeau04}, causing it to merge 
with its binary companion nearly independent of the semi-major axis or expansion factor.  This is 
the case in every NSC in our sample, with the exception of M33 due to its lower velocity dispersion.  
If the binary progenitors are destroyed, then Kozai-Lidov oscillations cannot operate.  
Therefore, we 
do not expect binary mergers due to Kozai-Lidov oscillations to have contributed significantly to any of the anomalous 
blue star populations currently observed in the inner regions of the (old) NSCs in our sample of galaxies.  

Although the mass segregation timescale for a massive population can approach the cluster age in the NSCs in our 
sample, heavy objects could still have been delivered via this mechanism to the inner nuclear regions in non-negligible 
numbers by the present-day (provided they are born at sufficiently small clustercentric radii).  BH-BH binary mergers 
induced by Kozai-Lidov oscillations with the central SMBH 
could still be relatively high at the present epoch relative to the rates of stellar binary mergers 
\citep[e.g.][]{vanlandingham16}.  However, these BH-BH binaries 
should either be primordial or have formed very far out in the NSC and only recently migrated inward.  This is because the rate 
of \textit{stellar} binary destruction is expected to be very high.  If few binaries exist at the present epoch, then BHs cannot 
be exchanged in to them via dynamical encounters to form BH-BH binaries.  Additionally, due to the presence of 
the massive central BH, BH binary formation via tidal capture can be unlikely since the timescale for this is comparable 
to the timescale on which the BH binary will experience a strong encounter with the central SMBH and itself 
be stripped of its companion \citep[e.g.][]{leigh14}.

\subsection{X-ray binaries and millisecond pulsars} \label{xrays}

In order to estimate the numbers of X-ray binaries using our derived collision rates (i.e. the number of dynamically formed 
X-ray binaries), we need to know the fractions of 
stellar remnants, namely white dwarfs f$_{\rm wd}$, neutron stars f$_{\rm ns}$ and black holes f$_{\rm bh}$.  Unfortunately, 
these numbers are highly uncertain, due to uncertainties in the initial stellar mass function at the high-mass end (which are 
significant), neutron star kicks at birth which can eject them from the nucleus, etc.  Regardless, we can make general 
comments about the expected frequencies of X-ray binaries in our sample of nuclei, which is also highly sensitive 
to the cluster binary fraction f$_{\rm b}$.  Based on our results for the 1+2 collision rates, we expect very few binaries 
to have survived until the present-day.  This is due both to the high 1+2 collision rates, but also to the separations 
corresponding to the hard-soft boundaries in such high velocity dispersion environments, which are near contact.  With 
this in mind, it seems more likely that the nuclei with the lowest velocity dispersions should harbor the largest 
binary fractions, namely M32 and M33, which have hard-soft boundaries of 0.04 AU and 0.6 AU, respectively.  In both 
M32 and M33, the 1+2 collision rates are also the highest, such that exchange encounters should occur very 
frequently.  Thus, we might expect a larger fraction of cataclysmic variables, NS and BH X-ray binaries as well as 
millisecond pulsars (MSPs) in the inner nuclear regions of M32 and especially M33, relative to the MW and 
M31 which are home to more hostile environments for long-term binary
survival in the inner $\lesssim$ 1 pc. 
With that said, \citet{leigh14} recently showed (which was subsequently confirmed by \citet{giersz15}) that the 
presence of a massive black hole near the cluster centre can 
efficiently destroy X-ray binaries, lowering their numbers significantly relative to what would be predicted in the 
absence of a massive BH.  Thus, putting this all together, we expect M33 to have the largest fraction 
of X-ray binaries per unit cluster mass, since it has both the lowest velocity dispersion and the smallest upper limit on 
the SMBH mass.  Indeed, constraining the formation of X-ray binaries in the M33 nucleus 
could help to place constraints on the possible presence of a central BH, which remains controversial and 
is at best, of very low mass (i.e. $\lesssim$ 1500 M$_{\rm \odot}$).  

Finally, we note that various processes not considered in our study may effectively replenish the nuclear regions 
of galaxies with binaries and compact objects, possibly enhancing the formation of X-ray binaries and MSPs in NSCs.
For example, nuclear stellar clusters may result from the continuous infall of star clusters that disrupt close to the 
SMBH \citep[e.g.,][]{2015ApJ...812...72A}.  Such stellar clusters may harbor MSPs, X-ray binaries as well as
large populations of stellar remnants.  In addition, \citet{2009ApJ...698.1330P} suggest that the disruption of triple stars 
could leave behind a binary in a close orbit around an SMBH; the rate of triple disruptions (brought in to the inner 
nuclear regions at late times either via NSC infall or mass segregation) could be high enough to 
serve as a continuous source of binaries close to the SMBH.

\section{Summary} \label{summary}

In this paper, we consider the origins of the enigmatic stellar populations observed in the 
nuclear star clusters of Local Group galaxies, specifically the Milky Way, M31, M32 and M33.  
These curious populations are blue stars found in the inner $\lesssim$ 0.1 pc of three out of 
the four galactic nuclei considered here.  The origins of these centrally-concentrated blue stars are not 
known.  Several candidates have been proposed in the literature, including blue straggler stars, extended horizontal 
branch stars and young recently formed stars.  Here, 
we calculate rough order-of-magnitude estimates for various rates of collisions, as a 
function of the host nuclear cluster environment and distance from the cluster centre.  We subsequently 
quantify the contributions to the enigmatic blue stars from BSs, extended horizontal branch stars 
and young recently formed stars.  

The collision rates are sufficiently high that blue stragglers, formed via direct collisions between single main-sequence stars, 
could contribute non-negligibly ($\sim$ 1-10\%) to the observed surface brightness profiles, in nearly (with the 
exception of the MW) every nucleus in our sample.  However, 
the radial profiles of the collision products do not always predict a strong central concentration of these objects, as 
in the MW (for our assumed density profile, which has a constant density core), M32 and, to a lesser extent, M31.  
In M31 and M32, the rates of MS+RGB collisions might only be sufficiently high for 
individual giants to undergo multiple collisions within $\lesssim$ 0.1 pc from the galaxy 
centre.  Only here could a paucity of giants be observed, since the results of stellar evolution models presented in 
this paper suggest that the envelope 
must be nearly fully stripped to destroy an RGB star and leave an EHB star in its place.  

The rates of collisions between white dwarfs and MS stars, which are expected to typically ablate 
the MS stars, are sufficiently high that this could offer a steady supply of gas to the inner nucleus and 
subsequently contribute non-negligibly to the total gas reservoir needed to seed star formation in every galactic 
nucleus in our sample.  This scenario seems the most likely 
(dominant?) of those considered in this paper for explaining the origins of the inner blue stars, relative to a 
direct collisional origin.  However, we suggest that the gas might be more likely to simply accrete onto 
pre-existing old stars rather than fragment and form new stars.  This could offer a new channel for BS 
formation or the rejuvenation of old MS stars, by creating "non-binary mass transfer blue stragglers" in the 
inner nucleus.  We emphasize, however, that more work needs to be done to better constrain the expected 
accretion rates in this scenario, which are poorly known.

Our results suggest that collisional processes could contribute significantly to the observed blue excesses in 
M31 and M33, via several different channels.  More sophisticated theoretical models are needed, however, in 
order to properly compliment existing and future observations.  
For example, accurate collision surface brightness profiles, which can be derived from theoretical collision 
number profiles (such as those presented in this paper), can be converted into radial colour profiles, multiplied 
by a suitable constant (to ensure not under- or over-subtracting) and subtracted from the observed 
radial colour profiles.  Any remaining colour gradient observed in the residuals should then be indicative 
of a real age gradient in any high-density unresolved stellar population.  We caution that this method is best 
suited to quantifying only the \textit{gradient} in an observed age spread, and not the relative sizes of the underlying blue 
and red populations.

\section*{Acknowledgments}

We would like to kindly thank Barry McKernan and Saavik Ford for useful discussions and suggestions.  
NL acknowledges the generous support of an AMNH Kabfleisch Postdoctoral Fellowship.  
FA acknowledges support from a NASA Fermi Grant NNX15AU69G and from a CIERA postdoctoral fellowship at Northwestern
University. Financial support was provided to NS by NASA through Einstein Postdoctoral Fellowship Award Number PF5-160145.  
FA and DM acknowledge insightful conversations with Eugene Vasiliev that stimulated them to conduct the 
calculation presented in Section~\ref{ehbform}.  Financial support was granted to DM by the National Science Foundation under 
grant no. AST 1211602 and by the National Aeronautics and Space Administration under grant no. NNX13AG92G.

\bsp

\label{lastpage}

\end{document}